\documentclass{article}
\usepackage{psfig}
\usepackage{emulateapj}

\def \nh {N${\rm _H}$}

\def \hcm {\hbox {\ifmmode $ atom cm$^{-2}\else atom cm$^{-2}$\fi}}
\def \arcmin {\hbox{$^\prime$}}

\def\approxgt{\mathrel{\hbox{\rlap{\lower.55ex \hbox {$\sim$}}
        \kern-.3em \raise.4ex \hbox{$>$}}}}
\def\approxlt{\mathrel{\hbox{\rlap{\lower.55ex \hbox {$\sim$}}
        \kern-.3em \raise.4ex \hbox{$<$}}}}

\slugcomment{ApJ, accepted Oct 28, 2002}

\begin{document}

\title{Soft X-ray excess emission in clusters of galaxies observed with XMM-Newton} 

\author{J. Nevalainen$^{1,2,3}$, R. Lieu$^{2}$, M. Bonamente$^{2}$, D. Lumb$^{3}$}
\affil{Harvard - Smithsonian Center for Astrophysics, Cambridge, USA$^{1}$\\ 
University of Alabama in Huntsville, Huntsville, USA$^{2}$\\
ESTEC, Noordwijk, Netherlands$^{3}$}

\begin{abstract}
We present results on the spectroscopic analysis of XMM-Newton EPIC data of the central 0.5 h$_{50}^{-1}$ Mpc
regions of the clusters of galaxies Coma, A1795 and A3112. The temperature
of the hot intracluster gas as determined by modeling the 
2 -- 7 keV PN and MOS data is consistent with that inferred from the FeXXV-FeXXVI line ratio.  
A significant warm emission component at a level above the systematic uncertainties is evident in the data
and confirmed by ROSAT PSPC data for Coma and A1795. 
The non-thermal origin of the phenomenon cannot be ruled
out at the current level of calibration accuracy, but the thermal
model fits the data significantly better, with temperatures 
in the range of 0.6 -- 1.3 keV and electron densities of the order of 
$10^{-4}$ -- $10^{-3}$ cm$^{-3}$. In the outer parts of the clusters
the properties of  the warm component are marginally consistent with
the results of recent cosmological simulations, which predict a large
fraction of the current epoch's bayons located in a warm-hot intergalactic
medium (WHIM). However, the derived densities are too high in the cluster
cores, compared to WHIM simulations, and thus more theoretical work is
needed to fully understand the origin of the observed soft X-ray
excess.
\end{abstract}


\section{Introduction}
In 1996, Lieu et al. (1996a,b) reported the discovery of excess EUV
and soft X-ray emission above the contribution from the hot ICM in the nearby Virgo and Coma
clusters using data from Extreme UltraViolet Explorer ({\it EUVE}) and ROSAT
PSPC.
Later, EUV excess emission was detected by {\it EUVE} in A1795 by Mittaz, Lieu
and Lockman (1998) and in A2199 by Lieu, Bonamente and Mittaz (1999a); 
these results were disputed by Bowyer, Berghoefer and Korpela (1999),
and further affirmed by  Lieu et al. (1999b; A2199) Bonamente, Lieu and Mittaz
(2001a; A1795 and Virgo). 
The {\it BeppoSAX} LECS instrument provided also a positive detection of
excess emission in A3571 (Bonamente et al. 2001c) and A2199 (Kaastra
et al. 1999), the latter challenged by Berghoefer and Bowyer (2002)
yet rebutted by Kaastra et al. (2002).
Analysis of {\it ROSAT} PSPC data has yielded further evidence for soft X-ray excess
emission in Coma (Arabadjis and Bregman, 1999), Virgo (Buote, 2001; Bonamente
Lieu and Mittaz, 2001b),  Shapley (Bonamente et al. 2001c) and  S\'{e}rsic 159-03 (Bonamente,
Lieu and Mittaz 2001d). Recently, a sample of clusters studied with
PSPC yielded significant soft excess emission in $\sim$50\% of the clusters (Bonamente et al. 2002).

Possible scenarios, involving thermal or non-thermal emission, for the
origin of the soft excess have been proposed since the discovery of
the phenomenon.
The thermal model requires large amounts of warm baryons 
(Lieu et al 1996a,b; Mittaz, Lieu \& Lockman 1998; 
see also Bonamente et al 2001b,c,d). They are currently 
believed to reside in warm-hot {\it inter-}cluster filaments, as recently modeled in large-scale hydrodynamical 
simulations (e.g., Cen et al. 1999,2001; Dave et al. 2001). In the non-thermal model the soft excess is due to
inverse-Compton interaction between the cosmic microwave background and a population of intracluster relativistic 
electrons (Hwang 1997, Sarazin \& Lieu 1998, Lieu et al. 1999).

The previously analyzed ROSAT PSPC data indicate that the excess emission is probably of thermal origin, although the 
limited spectral resolution of PSPC could not completely rule out the presence of non-thermal radiation 
(e.g., Bonamente et al. 2002). The superior spectral resolution, the large bandpass 
coverage (0.2 -- 10 keV) and the large collecting area of the XMM-Newton EPIC makes it well suited for studying
the soft excess phenomenon. With EPIC, one can for the first time constrain simultaneously the properties of the
hot gas and the soft component in clusters of galaxies. In this work we analyse the PN and MOS data from the 
central 0.5 h$_{50}^{-1}$ Mpc regions of clusters Coma, A1795 and A3112. 
We will present results on the detection and modeling of the soft component and
outline a possible scenario responsible for this
phenomenon. We will study the accuracy of the EPIC calibration by comparing our results with those of 
the reported EPIC calibration activities, and with the published ROSAT PSPC data. 

We consider uncertainties and significances at 90\% confidence level, and use
H = 50 $\times$ h$_{50}$ km s$^{-1}$ Mpc$^{-1}$, unless stated otherwise.

\begin{table*}[htb]
\begin{center}
\caption[]{Basic information on the observations}
\vspace{0.2cm} 
\begin{tabular}{llllllllllllll}
\hline
\multicolumn{14}{l}{\vspace{-0.2cm}} \\ 
cluster  & z & obs id & obs date & \multicolumn{3}{c}{livetime$^{a}$}
& \multicolumn{3}{c}{count rate cut$^{b}$} & \multicolumn{3}{c}{exposure$^{c}$} & filter \\
         &     &        & yyyy-mm-dd &  \multicolumn{3}{c}{ks} & \multicolumn{3}{c}{} &  \multicolumn{3}{c}{ks} & \\
         &     &        &            &  \multicolumn{3}{c}{\hrulefill} & \multicolumn{3}{c}{\hrulefill} &  \multicolumn{3}{c}{\hrulefill} & \\
         &     &        &            & PN & M1 & M2 &  PN & M1 & M2 &  PN & M1 & M2 & \\
\multicolumn{14}{l}{\vspace{-0.2cm}} \\
Coma     & 0.023 & 0153750101  & 2001-12-04 & 20 & 25 & 25 & 55 & 15 & 15 & 17 & 21 & 20 & medium \\
A1795    & 0.062 & 009782010   & 2000-06-26 & 42 & 48 & 48 & 50 & 15 & 25 & 23 & 29 & 32 & thin  \\
A3112    & 0.076 & 0105660101  & 2000-12-24 & 17 & 23 & 23 & 50 & 15 & 15 & 17 & 22 & 22 & medium \\
\multicolumn{14}{l}{\vspace{-0.2cm}} \\
\hline
\multicolumn{14}{l}{\vspace{-0.2cm}} \\
\multicolumn{14}{l}{a: central CCD, after pipeline processing} \\
\multicolumn{14}{l}{b: counts in 100s bin at 12 -- 14 keV (PN) and 10 -- 12 keV (MOS)} \\
\multicolumn{14}{l}{c: after filtering the flares} \\
\smallskip
\end{tabular}
\label{info.tab}
\end{center}
\end{table*}

\section{Data Analysis}
The data used in this work are taken from XMM-Newton public data archive (Coma and A1795) and from the Guaranteed 
Time program (A3112). We used the pipeline processed products of Coma
and A1795 and processed the A3112 data using epchain and emchain tools in SAS 5.2.0.
Among the clusters available to us at the time of this writing,  
only these three were bright enough for our analysis.
For the details of the observations of Coma and A1795 we refer to the
previous published analyses of some of these data (Coma: Arnaud et al. 2001a;
A1795: Arnaud et al. 2001b, Tamura et al. 2001). We report the basic information of the 
specific data used in our work in Table \ref{info.tab}. 
All the data are obtained in the full frame mode of the EPIC cameras.

In order to identify the periods of significant soft proton flares, we
extracted light curves of the cluster data
in 100s bins and for energies $\geq$10~keV, where the cluster contribution is negligible due to the low effective area 
(see e.g. Marty et al. 2002). If the count rate in a given time bin exceeds our limits (see Table \ref{info.tab}), we will
exclude that bin at the spectrum extraction stage. The light curves show that A3112 data are free of strong flares 
but have some short term spikes. There is a strong flare towards the end of the observation 
of Coma, excluding which reduces the useful exposure time by 20\%. There are several strong flares during the 
observation of A1795, reducing the useful exposure by 50\%.  After the flare 
filtering, there remains enough photons 
for our analysis in all clusters.

We extracted the cluster spectra using only photons designated with patterns 0 for PN and 0 -- 12 for MOS. 
Excising point sources and bad pixels, we obtained the spectra in two large radial bins, 
0-0.2-0.5 h$_{50}^{-1}$ Mpc, which correspond to 0-5.2\arcmin-13.0\arcmin\ for Coma, 
0-2.1\arcmin-5.1\arcmin\ for A1795 and 0-1.7\arcmin-4.3\arcmin\ for A3112. 
For the energy redistibution of PN, MOS1 and MOS2 we use the
calibration files epn\_ff20\_sY9.rmf , m1\_r5\_all\_15.rmf and m2\_r5\_all\_15.rmf ,
respectively. We created the  effective area files with arfgen-1.44.4
tool within SAS distribution, using the calibration information
available in March 2002.

For the background estimate, we used the blank sky data (Lumb et al. 2002) extracted at the same detector 
coordinates as the source spectra, using the same count rate criteria
in the $>$ 10 keV band as for the data.
We limit our spectral analysis to 0.3 -- 7.0 keV band to ensure that 
at low energies the source signal level is more than 10 times above that of the background and 5 times at the highest
energies. This is needed to avoid any errors in the background subtraction, as explained below.
We use the 12 -- 14 keV (PN) and 10 -- 12 keV (MOS) band of source and background data to normalize 
the background to correspond the background level during the cluster
observations. Here we effectively assume that any excess background
has the same spectral shape as the blank sky. While the spectra due to
the possible residual soft proton flares may be different from the
blank sky spetrum, the above assumption has no effect on our results: 
Owing to the adopted criteria for the ratio of source to background levels, the typical background variation by 
10 -- 20 \% leads to smaller than 5\% variations in the background
subtracted data, i.e. below the level of systematic uncertainties. 
The blank sky data were obtained with a thin filter, while Coma and A3112 observations were made with the medium 
filter. Although the effective area of PN for the thin filter is 50\% higher than that for the medium filter at
0.3 keV, the difference reduces to 1\% at 1 keV, then the small
background at low energies results in a maximum oversubtraction of less than
5\% at all energies.

We binned the spectra to achieve at least 20 counts per bin, and further required that the energy resolution 
(FWHM) is oversampled by less than a factor of 3.  We determined this resolution at different energies 
by using XSPEC to fold a narrow Gaussian at a given energy and measuring the energies where the 
redistributed counts drop below 0.5 of the peak value.  
At $<$ 0.5 keV this is not feasible due to strong deviation of the energy re-distribution function from the 
gaussian shape, i.e. even if the FWHM was small, 
the broad shoulders distribute the counts much further. We thus used the FWHM values found at 0.5 keV for lower 
energies. The resulting FWHM values range from 90 eV to 160 eV between energies
0.5 keV and 7 keV. 

\begin{table*}[htb]
\begin{center}
\caption[]{Comparison of different temperature measurements. ``Mekal'' refers to fitting 2 - 7 keV band with 
absorbed mekal model (repeated from Table \ref{soft_hard.tab}), ``cont'' corresponds to fitting the 2 - 6 keV energy band with absorbed bremsstrahlung
model, and ``Fe'' to the FeXXV -- FeXXVI line ratio measurement}
\vspace{0.2cm} 
\begin{tabular}{llllll}
\tableline
\multicolumn{6}{l}{\vspace{-0.2cm}} \\
                          & \multicolumn{3}{c}{PN}         & MOS & PN+MOS \\
                          & \multicolumn{3}{c}{\hrulefill} &    & \\ 
radii                     &  T$_{Fe}$        & T$_{cont}$           & T$_{mekal}$             &   T$_{mekal}$  & T$_{mekal}$     \\ 
                          &   (keV)             &  (keV)               & (keV)                & (keV)          &  (keV)    \\    
\multicolumn{6}{l}{\vspace{-0.2cm}} \\
\multicolumn{6}{c}{\bf Coma}            \\
0\arcmin -- 5\arcmin      & 8.7$^{+2.1}_{-1.4}$ & 9.2$^{+0.9}_{-0.6}$  & 9.2$^{+0.7}_{-0.6}$  & 9.9$^{+0.6}_{-0.5}$ & 9.6$^{+0.4}_{-0.3}$ \\
5\arcmin -- 13\arcmin     & 7.7$^{+1.0}_{-1.0}$ & 9.5$^{+0.8}_{-0.6}$  & 9.2$^{+0.6}_{-0.5}$  & 9.8$^{+0.6}_{-0.4}$ & 9.6$^{+0.3}_{-0.3}$\\
\multicolumn{6}{c}{\bf A1795}           \\              
0\arcmin -- 2\arcmin      & 5.4$^{+0.6}_{-0.6}$ & 5.2$^{+0.3}_{-0.2}$  & 5.4$^{+0.2}_{-0.3}$  & 5.5$^{+0.3}_{-0.2}$ & 5.5$^{+0.1}_{-0.1}$  \\
2\arcmin -- 5\arcmin      & 6.0$^{+0.7}_{-1.0}$ & 6.6$^{+0.5}_{-0.5}$  & 6.7$^{+0.4}_{-0.4}$  & 6.9$^{+0.4}_{-0.3}$ & 6.8$^{+0.1}_{-0.2}$\\
\multicolumn{6}{c}{\bf A3112}           \\
0\arcmin -- 1.5\arcmin    & ...                 & 4.3$^{+0.3}_{-0.2}$  &  4.6$^{+0.3}_{-0.2}$ & 4.9$^{+0.3}_{-0.2}$ & 4.8$^{+0.2}_{-0.1}$  \\
1.5\arcmin --  4.5\arcmin & ...                 & 5.2$^{+0.6}_{-0.5}$  & 5.4$^{+0.5}_{-0.4}$  & 5.1$^{+0.4}_{-0.4}$ & 5.2$^{+0.3}_{-0.2}$ \\
\tableline
\multicolumn{6}{l}{\vspace{-0.2cm}} \\
\label{T.tab}
\end{tabular}
\end{center}
\end{table*}

\section{Iron line ratio as thermometer}
\label{liner}
Because of the possible problems with the calibration of the spectral
response of PN and MOS, we first seek a way of characterizing the hot component 
without relying on the continuum. Such a tool is the temperature-dependent flux ratio of the emission lines due to
FeK$\alpha$ transition:  iron XXV (helium-like) and XXVI
(hydrogen-like). These lines cover a very narrow band ($\sim$ 300 eV)
at 6 - 7 keV at these redshifts and thus are quite insensitive to 
the details of the effective area function. The energy resolution of
PN ($\sim$ 150 eV FWHM) at 6 keV is adequate to resolve these lines
separately. Also, the number of the line emission photons in Coma and
A1795 ($\sim$1000) renders the ratio measurable in these clusters.

\subsection{Model}
To obtain a model for the line flux ratio as a function of temperature, 
we used the XSPEC model MEKAL and PN responses to simulate spectra for a
grid of temperatures with a step size of 0.1 keV, keeping
metal abundance at 0.3 of solar value, and normalization fixed
to unity in XSPEC units. Since the fluxes of the FeXXV and
FeXXVI lines have the same dependence on the metal abundance, as well
as on the overall model normalization (emission measure), the flux ratio is independent
on these parameters. 
Moreover, the exposure time used in the simulations was large enough to ensure negligible statistical 
errors. 

\begin{figure*}
\hbox{
\psfig{figure=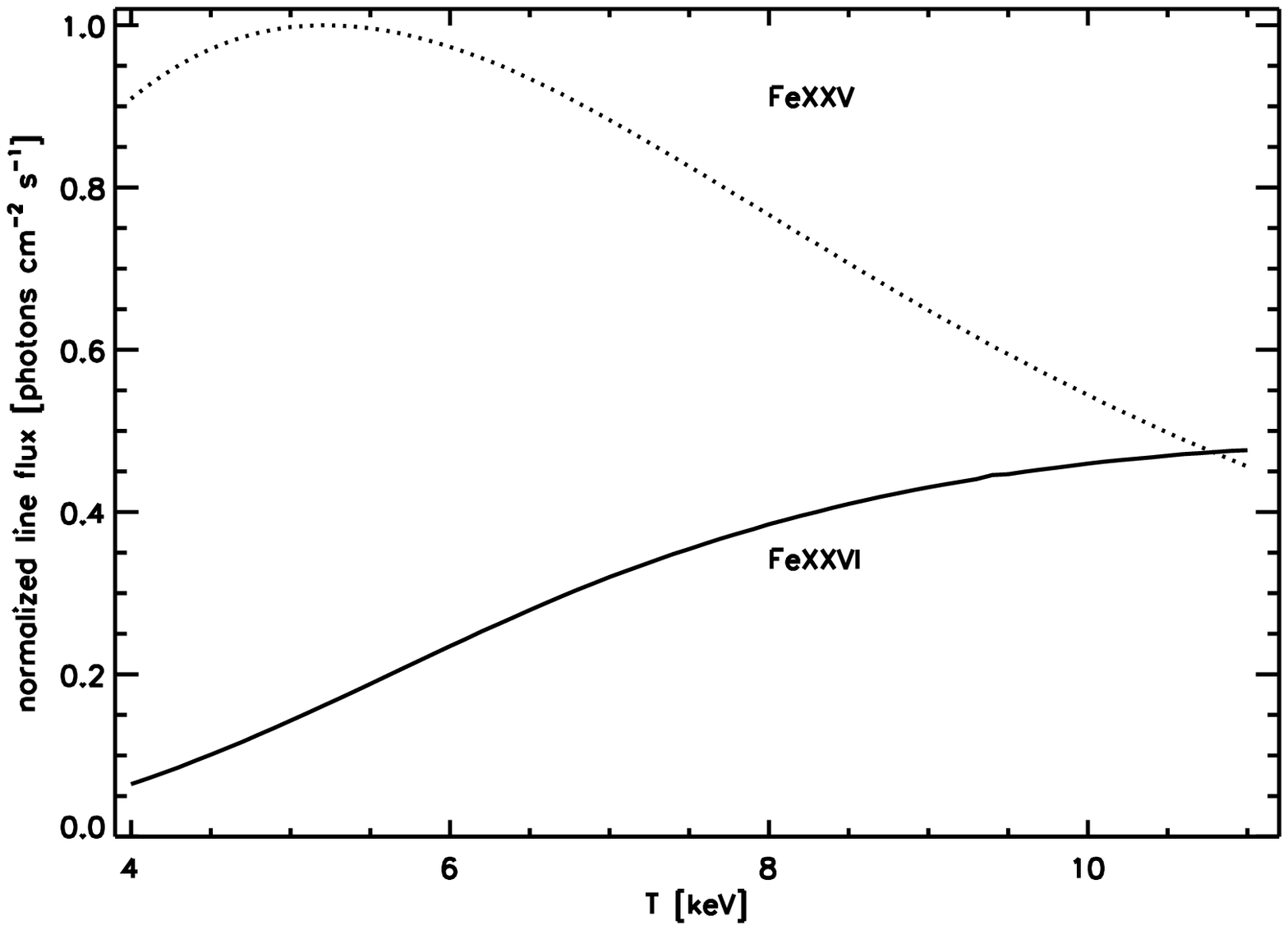,width=8.0cm,angle=0}
\psfig{figure=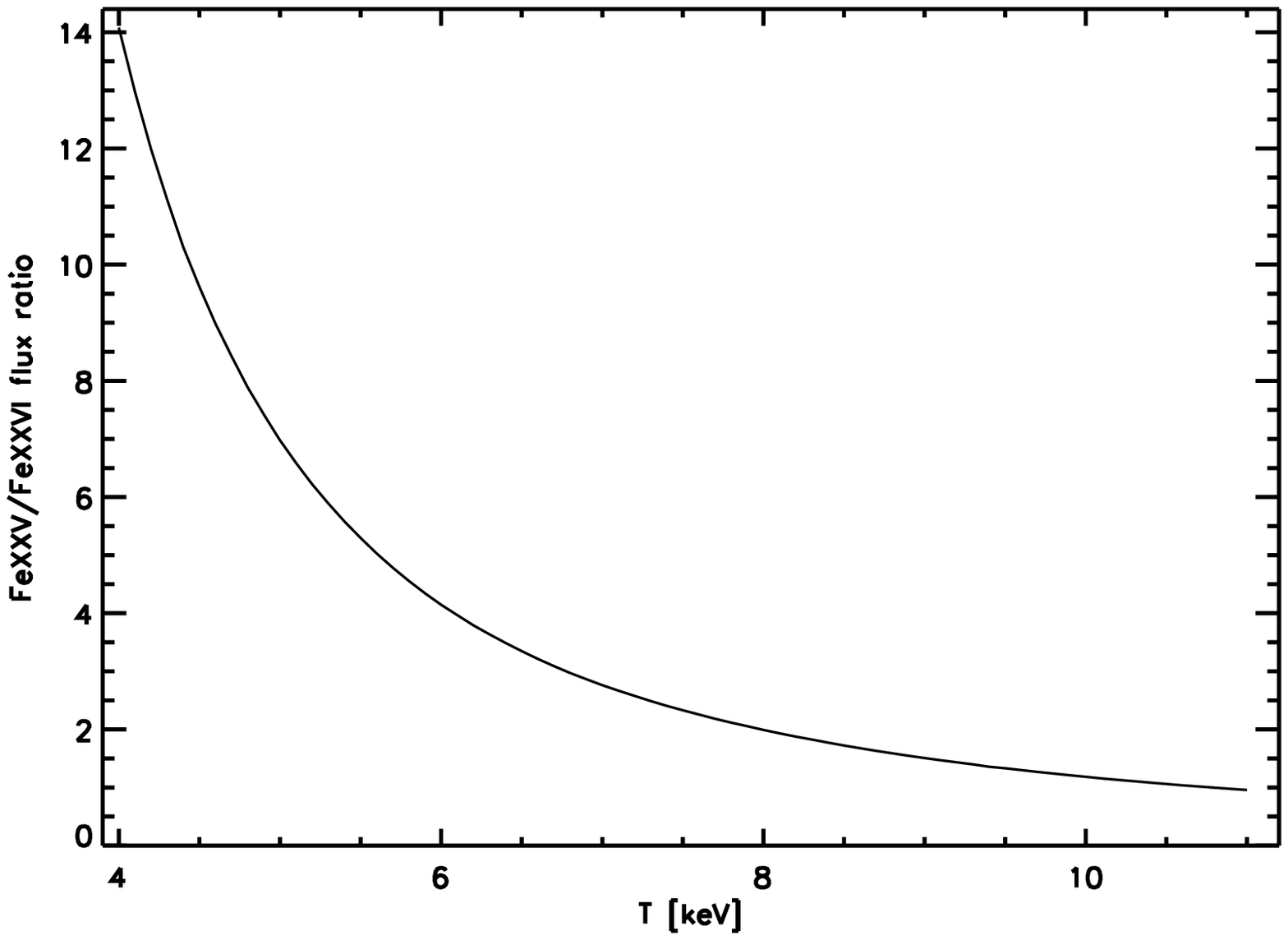,width=8.0cm,angle=0}}
\vspace{0.5cm}
\caption{The fluxes of FeXXV and FeXXVI emission lines in the mekal
model (scaled to the maximum flux of FeXXV) (left panel) and their ratio
(right panel) as a function of the electron temperature .}
\label{liner_mod.fig}
\end{figure*}

We then modeled the continuum of the simulated spectra with a
bremsstrahlung model, fixing the temperature to that used for the
simulation, and fitting the normalization using the data in energy
intervals of 5 -- 6 and 9 -- 10.5 keV. We then modeled the data in 6
-- 7.5 keV band using the bremsstrahlung model fixed as above and adding 
two Gaussians for the FeXXV and FeXXVI lines. We used the
best fit models to obtain the fluxes of the Gaussians (in photons cm$^{-2}$ s$^{-1}$)
and consequently determined the theoretical flux ratio of these two 
emission lines as a function of temperature (see Fig.\ref{liner_mod.fig}).
The change in relative abundance of FeXXV and FeXXVI ions with
temperature is reflected in the above curves, resulting in  
a decreasing flux ratio of FeXXV-to-FeXXVI emission lines
with increasing temperature, from value of 14 at 4 keV  to unity at
11 keV. At low temperatures the model ratio is very sensitive to the 
variation in temperature (between 4.0 and 4.5 keV the ratio chages
from 14 to 10) while it is less sensitive at high temperatures
(between 10.5 and 11.0 keV the ratio changes from 1.1 to 1.0).

\begin{table*}[htb]
\begin{center}
\caption[]{Results of single temperature MEKAL fit to 0.3 -- 7.0 keV PN data using 0\% or 5\% systematic errors and 
various values of \nh\
The statistical errors of \nh\ is $\sim$ $1
\times$ $10^{19}$ atoms cm$^{-2}$. The $\chi^{2}$ and the 90\% confidence interval of \nh\ are shown when \nh\ is treated as a free 
parameter}
\vspace{0.2cm} 
\begin{tabular}{llllllll}
\hline
\multicolumn{8}{l}{\vspace{-0.2cm}} \\ 
      &        & 0\%  & 5\%   & \multicolumn{2}{c}{5\%, \nh\ free} & & \\ 
\multicolumn{8}{l}{\vspace{-0.2cm}} \\
radii & d.o.f. & $\frac{\chi^{2}}{d.o.f.}$ & $\frac{\chi^{2}}{d.o.f.}$ & $\frac{\chi^{2}}{d.o.f.}$  & \nh$^{a}$ & \nh$_{G}$$^{b}$ & \nh$_{G}$$^{c}$ \\
\multicolumn{8}{l}{\vspace{-0.2cm}} \\
\multicolumn{8}{c}{{\bf Coma}}   \\
0\arcmin -- 5\arcmin      & 176 & 3.46 & 0.93 & 0.77  & 0$^{+0.1}_{....}$    & 0.9  & 0.9 \\
5\arcmin -- 13\arcmin     & 176 & 6.72 & 1.11 & 0.90  & 0$^{+0.1}_{....}$    & 0.9   & 0.9 \\
\multicolumn{8}{c}{{\bf A1795}} \\
0\arcmin -- 2\arcmin      & 177 & 4.04 & 1.11  & 0.94 & 0$^{+0.2}_{....}$ & 1.0  &  1.2 \\
2\arcmin -- 5\arcmin      & 176 & 2.76 & 1.04  & 0.91 & 0$^{+0.3}_{....}$ & 1.0  &  1.2 \\
\multicolumn{8}{c}{{\bf 3112}} \\
0\arcmin -- 1.5\arcmin    & 177 & 3.32 & 1.48  & 1.06 & 0.2$^{+0.5}_{-0.2}$ & 2.5    &   \\
1.5\arcmin --  4.5\arcmin & 177 & 1.90 & 1.28  & 1.04 & 0.7$^{+0.5}_{-0.6}$ & 2.5   &    \\
\tableline
\multicolumn{8}{l}{\vspace{-0.2cm}} \\
\multicolumn{8}{l}{a: \nh\ as a free parameter, in $10^{20}$ atoms cm$^{-2}$} \\
\multicolumn{8}{l}{b: fine beam 21 cm \nh, in $10^{20}$ atoms cm$^{-2}$} (Murphy et al. in prep.) \\
\multicolumn{8}{l}{c: broad beam 21 cm \nh, in $10^{20}$ atoms cm$^{-2}$} (Dickey \& Lockman 1990)) \\
\smallskip
\end{tabular}
\label{system.tab}
\end{center}
\end{table*}

\subsection{Results}
Next, we measured the line flux ratio in the cluster PN data by fitting the 6.0 --
7.5 keV (5.5 -- 7.0 keV) band for Coma (A1795) with an absorbed emission model
consisting of bremsstrahlung component and two Gaussians.
We chose to use these narrow bands instead of the full spectra in
order to minimize the dependence on calibration accuracy. Indeed,
using 2.0 -- 7.0 keV band for the fit leads to slightly 
larger FeXXV-to-FeXXVI ratio, but the corresponding variation in
temperature is negligible. Furthermore, using only the narrow band
around the iron lines results in a best fit that better describes the
line shapes, because of the higher relative importance of the line
photons in the $\chi^{2}$ sum. 

We defined the FeXXV line flux as FeXXVI line flux multiplied by
a constant and allowed this constant to vary, together with the FeXXVI
normalization, thereby giving the line flux ratio directly. We let
also all the bremsstrahlung parameters vary, as well as the gaussian widths and line centroids.
In the error analysis, the above parametrization of the line ratio
takes properly into account parameter correlations. 
When searching for the best fit, line centroids were allowed to vary due to possible gain calibration
inaccuracies, but the best fit values are virtually the same as
predicted. In the error analysis of the spectrum in  2 -- 5\arcmin\ region
of A1795, the model has too much freedom compared to
the quality of the data, resulting in a complicated $\chi^{2}$
distribution. Thus, for the line ratio error analysis, we fixed the 
line energies to the best fit values for all spectra, but this had
only effect in the outer part of A1795.

\begin{figure*}
\vbox{
\hbox{
\psfig{figure=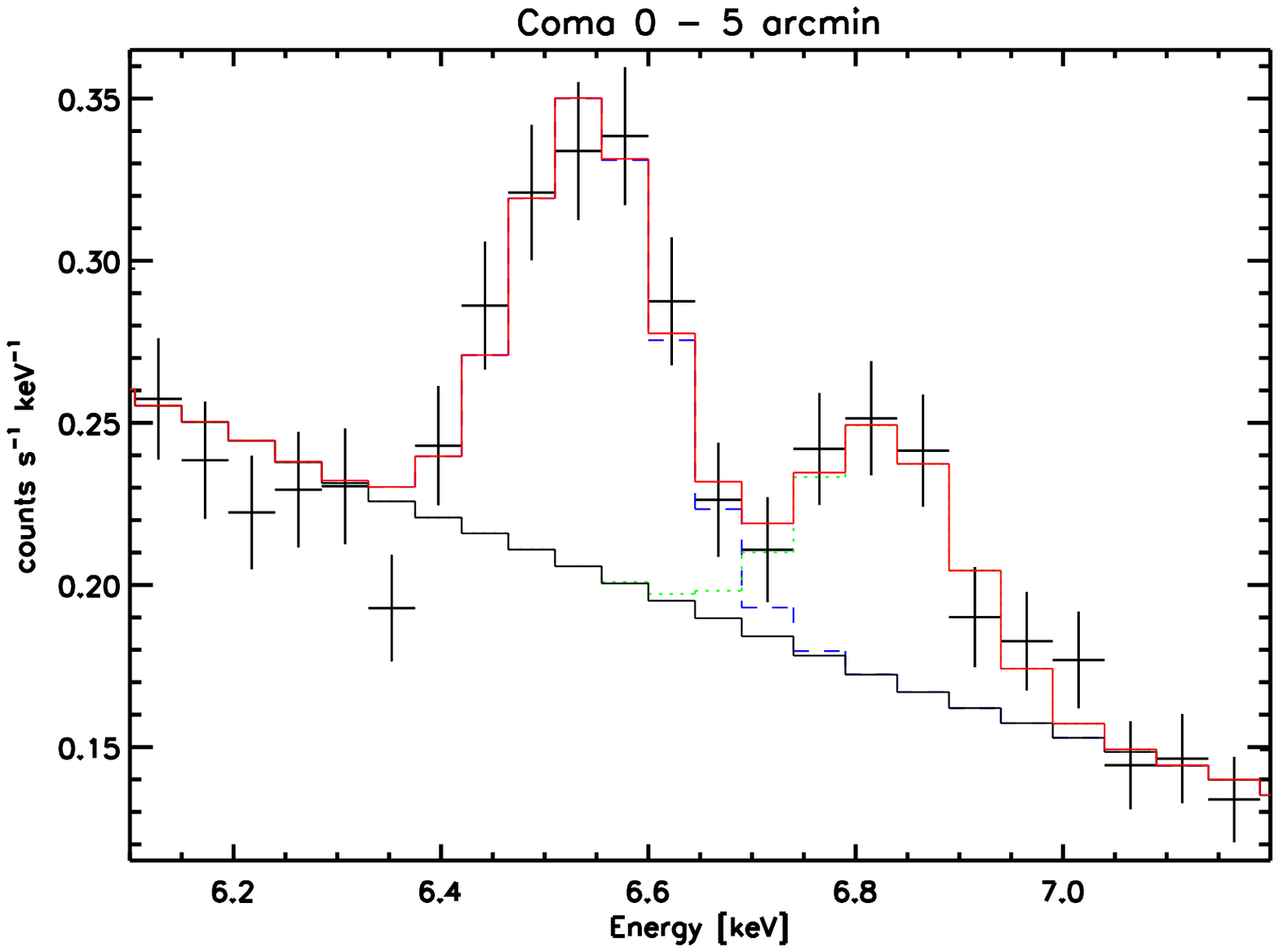,width=8.0cm,angle=0}
\psfig{figure=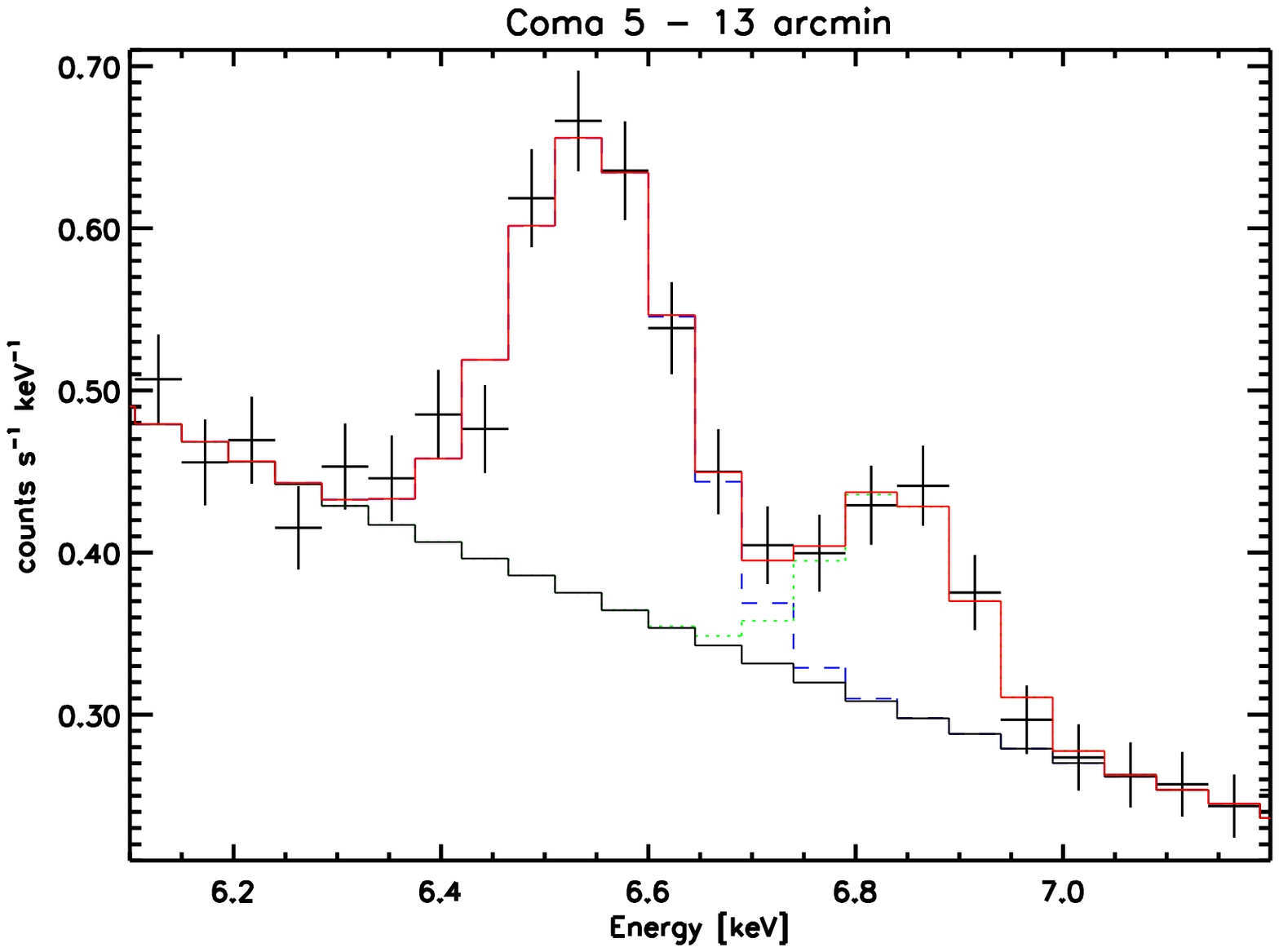,width=8.0cm,angle=0}}
\vspace{0.5cm}
\hbox{
\psfig{figure=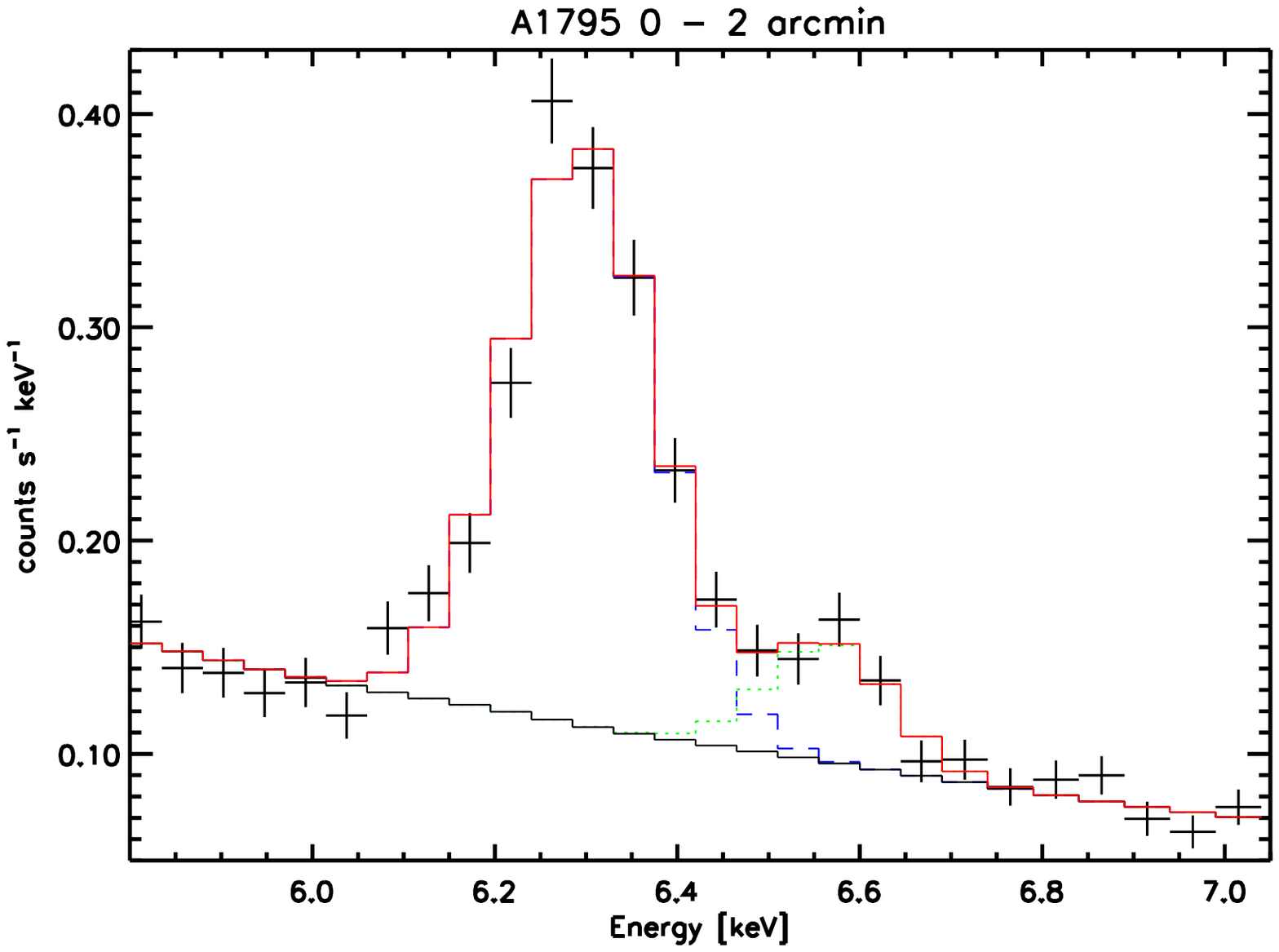,width=8.0cm,angle=0}
\psfig{figure=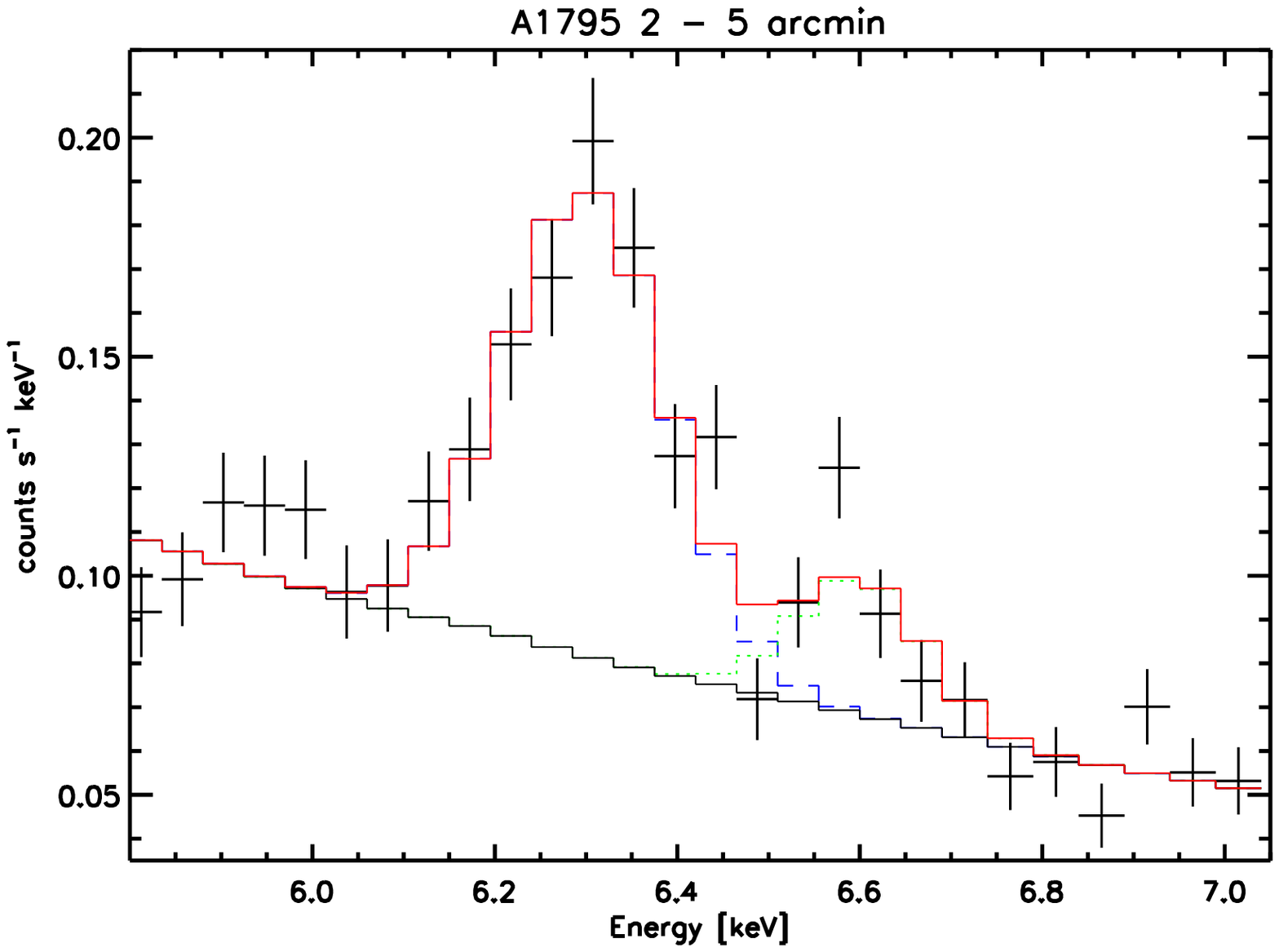,width=8.0cm,angle=0}}}
\vspace{0.5cm}
\caption{The PN spectra of Coma and A1795 around the FeXXV and FeXXVI
lines with 1 $\sigma$ uncertainties shown as crosses. The 
folded models for the continuum with and without the Gaussians added
are shown as solid lines. The dashed (blue) and the dotted (green)
lines show the FeXXV and FeXXVI lines separately.}
\label{liner_data.fig}
\end{figure*}

The best fit models are shown in Fig. \ref{liner_data.fig} and the
corresponding temperature values are reported in Table \ref{T.tab}.
At 90\% confidence the constraints of the flux ratio 
for Coma within radii 0--5--13\arcmin\ are 1.6$^{+0.9}_{-0.6}$ and 2.2$^{+0.9}_{-0.6}$
and for A1795  within radii 0--2--5\arcmin\ they are 5.6$^{+2.4}_{-1.7}$ and
4.2$^{+2.7}_{-1.2}$. The relative uncertainties of the line flux ratio
range between 30--70\% without no trend as a function of temperature,
since the strengths of the two lines behave in opposing ways when
varying the temperature. 
However, as seen above, at lower temperatures
the model is much more sensitive to temperature, i.e. at low
temperatures a similar variation in line ratio
corresponds to smaller variation in temperature than at high
temperatures. Thus, similar observable constraints in line ratio lead
to better constraints in the model temperature in A1795, compared with Coma.
At the lowest temperatures (4 keV) the FeXXVI line is weak 
and in case of A3112 it reaches the level of noise and thus the flux 
ratio of A3112 is uninformative. The resulting temperature values are
discussed further in Section \ref{isot}.

\section{Problems with isothermal modeling}
\label{isot}
\begin{figure*}
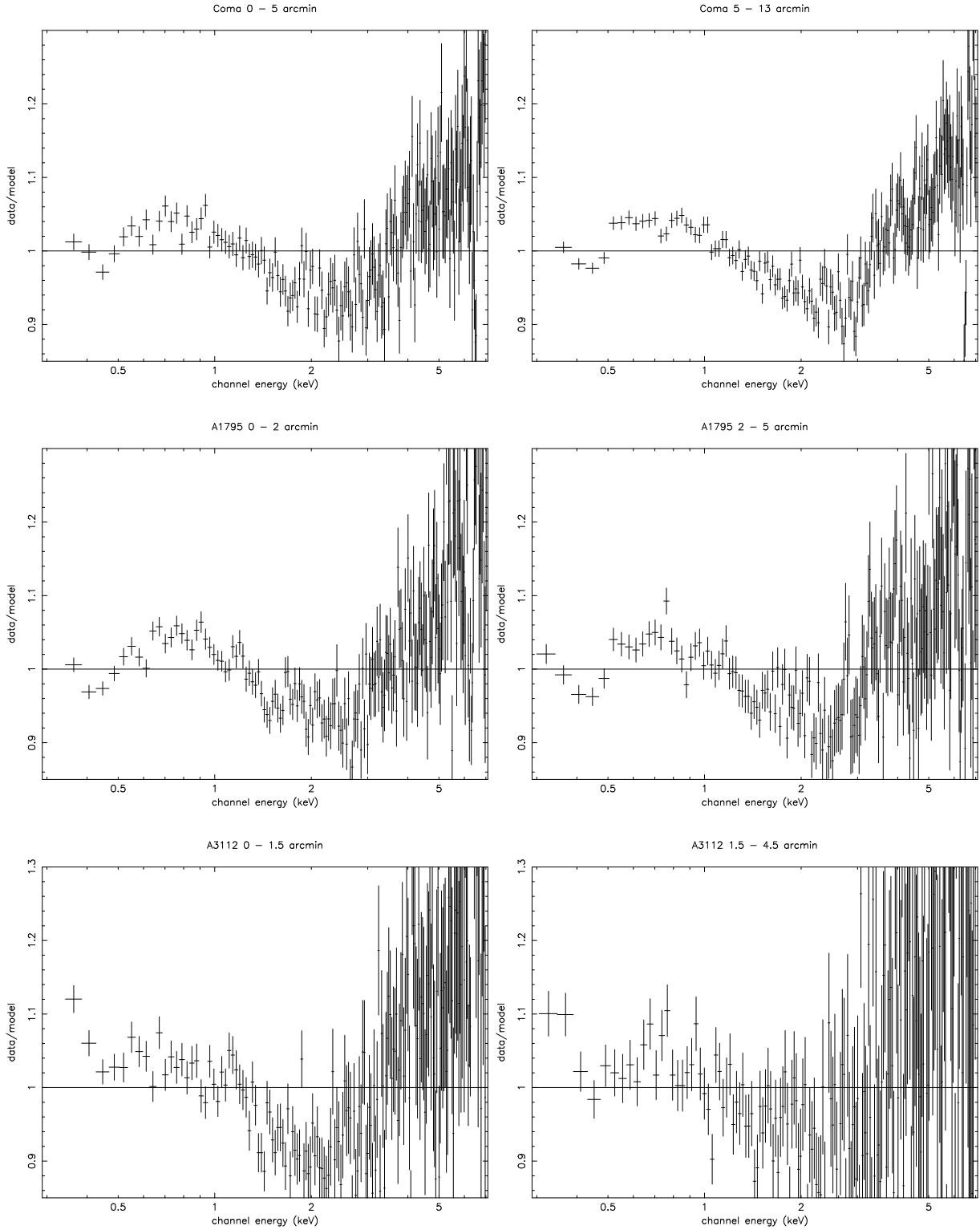

\vbox{
\hbox{
\psfig{figure=f3a.ps,width=8.0cm,angle=-90}
\psfig{figure=f3b.ps,width=8.0cm,angle=-90}}
\vspace{0.5cm}
\hbox{
\psfig{figure=f3c.ps,width=8.0cm,angle=-90}
\psfig{figure=f3d.ps,width=8.0cm,angle=-90}}
\vspace{0.5cm}
\hbox{
\psfig{figure=f3e.ps,width=8.0cm,angle=-90}
\psfig{figure=f3f.ps,width=8.0cm,angle=-90}}}
\caption{The ratio of the PN data of Coma, A1795 and A3112 to the best fit single temperature model to the 
0.3 - 7.0 keV band. When modeling the data no systematic errors were added.} 
\label{coma_ratio.fig}
\end{figure*}

We then modeled the whole 0.3 - 7.0 keV energy band data of PN with a single MEKAL model
(Mewe et al, 1995) absorbed by 
an HI column density (\nh) obtained with a fine beam aperture (20\arcmin\  , 
Murphy et al. in prep.), except for A3112, for which 
only the wide beam (1$^{\circ}$) value is available. 
We modeled the data without systematic errors, resulting in unacceptable fits
(see Table \ref{system.tab} and Fig. \ref{coma_ratio.fig}).
Between 0.5 and 1.0 keV there is a systematic 
positive deviation of the data from model at the 5 -- 10 \% level.
Between 1 and 4 keV there are negative residuals at the 10\% level and above 4 keV the 
positive residuals increase monotonically, reaching 20\% at 7 keV. Such large excursions are not 
consistent with the current understanding of the PN calibration (e.g. Briel, 2001; Kirsch, 2002).
Thus the 10\% and 20\% residuals, respectively at 2 -- 3 keV and above 5 keV disagree strongly with 
the reported residuals of a few \% at 2 -- 7 keV band using bright point sources with power-law spectra.
Addition of 5\% systematic errors (see e.g. Griffiths et al. 2002; Kirsch 2002; Snowden et al.
2002) to the whole band leads to improvement of the fits, 
but yet still not to an acceptable level for all regions in the 3 clusters (see Table \ref{system.tab}). 
Thus, at least one of the components involved in the above analysis,
i.e.  isothermal model, \nh\ from radio measurements or the adopted calibration information,  
must be wrong.

\subsection{Galactic absorption}
\label{nh}
A possible source of the above residuals is application of the incorrect absorption to the isothermal emission model.
To test this scenario, we allowed the \nh\ to vary as a free parameter and obtained statistically acceptable fits to PN 
data. However, the resulting \nh\
values are consistent with zero and significantly below the HI column
densities measured by narrow beam (20\arcmin\ ) radio 
observations at 21 cm wavelengths,
where the typical uncertainty is $\sim$ $ 1 \times 10^{19}$ cm$^{-2}$ 
(Murphy et al. in prep.). This discrepancy could be explained by 
assuming that towards the central 13\arcmin\ (Coma) and 5\arcmin\
(A1795 and A3112) cluster regions studied in our work, the Galactic \nh\ is 
by 0.5 - 2.0 $\times 10^{20}$ atoms cm$^{-2}$ smaller than 
that within the central 20\arcmin\ region covered by the radio measurements. 
But such \nh\ depletion in Galaxy, at a direction of these clusters,
would be a non-physical co-incidence and thus this explanation fails.
Moreover, the \nh\ values using broad beam ($\sim$1$^{\circ}$, Dickey and
Lockman 1990) and narrow beam (~20', Murphy et al.) 21 cm data for Coma and A1795 are
consistent (Table \ref{system.tab}). This further indicates that the galactic HI
column density is smoothly distributed in the direction of our clusters,
and that the application of \nh\ as measured in radio, is accurate. 

\begin{table*}[htb]
\begin{center}
\caption[]{The results of best fits to PN and MOS data to energy bands of 0.3 -- 2.0 keV and 2.0 -- 7.0 keV 
with mekal model. In the fits, we 5\% systematic errors and Galactic \nh\ are used}
\vspace{0.2cm} 
\begin{tabular}{lllllll}
\tableline
\multicolumn{7}{l}{\vspace{-0.2cm}} \\
      &   \multicolumn{2}{c}{0.3 -- 2.0 keV} &  \multicolumn{4}{c}{2.0 - 7.0 keV} \\
      & \multicolumn{2}{c}{\hrulefill}       & \multicolumn{4}{c}{\hrulefill}    \\
      &   PN  & MOS   & \multicolumn{2}{c}{PN}  &  \multicolumn{2}{c}{MOS} \\
      & T     & T     & T     & abund   & T     & ab    \\
      & [keV] & [keV] & [keV] & [Solar] & [keV] & [Solar] \\
\multicolumn{7}{l}{\vspace{-0.2cm}} \\
\multicolumn{7}{c}{{\bf Coma}} \\
0\arcmin -- 5\arcmin      &  4.9$^{+0.7}_{-0.4}$ & 6.9$^{+0.8}_{-0.9}$ & 9.2$^{+0.7}_{-0.6}$ &  0.21$^{+0.04}_{-0.04}$ & 9.9$^{+0.6}_{-0.5}$ &  0.23$^{+0.03}_{-0.03}$ \\
5\arcmin -- 13\arcmin     &  4.8$^{+0.5}_{-0.4}$ & 6.5$^{+0.8}_{-0.7}$ & 9.2$^{+0.6}_{-0.5}$ & 0.20$^{+0.04}_{-0.03}$  & 9.8$^{+0.6}_{-0.4}$ &  0.24$^{+0.03}_{-0.02}$ \\
\multicolumn{7}{c}{{\bf A1795}} \\
0\arcmin -- 2\arcmin      & 3.4$^{+0.2}_{-0.2}$ & 4.6$^{+0.4}_{-0.3}$   & 5.4$^{+0.2}_{-0.3}$ & 0.39$^{+0.04}_{-0.04}$  & 5.5$^{+0.3}_{-0.2}$ &  0.42$^{+0.03}_{-0.03}$ \\
2\arcmin -- 5\arcmin      & 4.1$^{+0.4}_{-0.3}$ & 5.5$^{+0.5}_{-0.4}$   & 6.7$^{+0.4}_{-0.4}$ & 0.27$^{+0.04}_{-0.04}$   & 6.9$^{+0.4}_{-0.3}$ &  0.29$^{+0.03}_{-0.03}$ \\
\multicolumn{7}{c}{{\bf A3112}} \\
0\arcmin -- 1.5\arcmin    & 2.6$^{+0.2}_{-0.1}$ &  3.5$^{+0.3}_{-0.2}$ & 4.6$^{+0.3}_{-0.2}$ & 0.50$^{+0.07}_{-0.06}$ & 4.9$^{+0.3}_{-0.2}$ &  0.47$^{+0.04}_{-0.04}$ \\ 
1.5\arcmin --  4.5\arcmin & 3.0$^{+0.3}_{-0.2}$ &  3.4$^{+0.3}_{-0.2}$ & 5.4$^{+0.6}_{-0.4}$ & 0.28$^{+0.07}_{-0.06}$ & 5.1$^{+0.4}_{-0.4}$ &  0.34$^{+0.06}_{-0.05}$ \\
\tableline
\multicolumn{7}{l}{\vspace{-0.2cm}} \\
\end{tabular}
\label{soft_hard.tab}
\end{center}
\end{table*}

\subsection{Calibration issues}
\label{calib}
\subsubsection{2 -- 7 keV band}
\label{hard}
To study the accuracy of the high energy band calibration, 
we fitted the 2 -- 7 keV energy band data in PN and MOS separately,
using MEKAL model and including 5\%
systematic errors in the fits. The resulting fits are acceptable, 
and the best-fit values of temperatures and metal abundances (see Table \ref{soft_hard.tab}) in PN and MOS
are in good agreement. This implies that either the hard band EPIC
calibration is accurate or that both PN and MOS are miscalibrated in a similar
manner.

The hard band temperatures of Coma and A1795 are 
consistent with those obtained using the 2 -- 10 keV band data from
BeppoSAX (deGrandi \& Molendi, 2002).
This indicates that there is no significant hard band miscalibration in PN and MOS.
Furthermore, in A1795, and in the center of Coma, the line ratio
analysis (Section \ref{liner}) yields temperatures consistent with those of the continuum fit and the MEKAL 
fit to 2 -- 7 keV PN and MOS data. 
These agreements imply that the hard band calibration of PN and MOS is
accurate within the level of the statistical uncertainties in our
work. 
In particular there is no evidence for any systematic trend caused by the 2 -- 7 keV 
miscalibration that biases our application of the MEKAL model to the high energy data.
In the outer part of Coma, the continuum temperatues are somewhat 
higher than the line ratio values.
The difference in Coma may be due to non-thermal emission as observed at hard X-ray band with BeppoSAX PDS 
(Fusco-Femiano et al. 1999). Extrapolating the best-fit power-law spectrum of PDS 20 -- 100 keV data to 
PN energies, and adding the expected PN flux from the hot gas has the effect of flattening 
the 2--7 keV slope, while leaving the line ratio unchanged. 
For Coma, we will compare the outcome of using either the line ratio temperatures or the MEKAL fit
values in Section 5.1.

\subsubsection{0.3 -- 2.0 keV band}
In order to assess the low energy band calibration accuracy, 
we repeated the isothermal analysis in the 0.3 -- 2.0 keV energy band.
In case of purely isothermal emission, the application of accurate absorption model
and accurate high energy band calibration information (as justified above),
should yield consistent temperatures in both bands for a given cluster
and a given region.
However, while the resulting fits to low energy band data are
statistically acceptable, they are significantly and systematically different from 
those obtained in the high energy band (see Table
\ref{soft_hard.tab}). The temperatures derived in
the low energy band using PN (MOS) are 2 -- 4 keV ( 1 -- 3 keV) smaller  
than those obtained using the high energy band. 
These conclusions do not change whether or not we include the cooling
flow regions of A1795 and A3112 in the comparison, i.e. the cooling does not explain
these discrepancies.
The significant underestimation of the low energy effective areas of
PN and MOS, and consequent underprediction of the thermal model
would lead to the softening of the best-fit spectra at low energies,
or, if the model is kept fixed to that derived using the high energy
band data, to soft excess. 
This would also explain why the fitted \nh\ values are much smaller
than the radio measurements (see Section \ref{nh}). 

However, the published XMM-Newton calibration works on power-law sources like BL-LAC objects 
(e.g. Briel et al. 2002, Ferrando et al. 2002, Haberl et al. 2002) 
do not report soft excesses or sub-galactic \nh. Also, analysis of PN
data of a large QSO sample (Akylas et al. 2002) with Galactic \nh\
yields no soft excess in these objects. Thus it seems likely that
a significant miscalibration is not the reason for the soft excess we
observe here. Rather, it is implied that the assumption of
isothermality is wrong and that the soft excess is real, as we assume in the following.

\begin{table*}[htb]
\begin{center}
\caption[]{Properties of the thermal soft component. Luminosities are
obtained in the 0.2 - 2.0 keV band and are in units of 10$^{43}$ erg
s$^{-1}$, using H = 50 km s$^{-1}$ Mpc$^{-1}$.} 
\vspace{0.2cm}
\begin{tabular}{lllllllll}
\tableline
\multicolumn{9}{l}{\vspace{-0.2cm}} \\
      & \multicolumn{2}{c}{PN}        & \multicolumn{2}{c}{MOS} & \multicolumn{4}{c}{PN + MOS} \\  
      & \multicolumn{2}{c}{\hrulefill} & \multicolumn{2}{c}{\hrulefill} & \multicolumn{4}{c}{\hrulefill} \\  
radii & T            &      ab        &      T       &  ab      & T & ab & L$_{warm}$ &  $\frac{L_{warm}}{L_{hot}}$   \\
      & [keV]        &   [solar]      &  [keV]       &  [solar] &  [keV]       &  [solar] &   &  \\ 
      &              &                &              &          &            &  \\
\multicolumn{9}{c}{\bf Coma}            \\
0\arcmin -- 5\arcmin       & 0.84$^{+0.16}_{-0.10}$ & 0.03$^{+0.03}_{-0.02}$  & 0.94$^{+0.14}_{-0.12}$ & 0.00$^{+0.02}_{-...}$  & 0.89$^{+0.19}_{-0.15}$ & 0.02$^{+0.05}_{-0.02}$ & 1.1$^{+0.3}_{-0.2}$ & 0.14$^{+0.03}_{-0.03}$ \\
5\arcmin -- 13\arcmin      & 0.82$^{+0.13}_{-0.08}$ & 0.03$^{+0.03}_{-0.02}$  & 0.90$^{+0.11}_{-0.10}$ & 0.00$^{+0.02}_{-...}$  & 0.86$^{+0.15}_{-0.12}$ & 0.02$^{+0.04}_{-0.02}$ & 3.1$^{+0.6}_{-0.6}$ & 0.15$^{+0.03}_{-0.03}$ \\
\multicolumn{9}{c}{\bf A1795}           \\              
0\arcmin -- 2\arcmin       & 0.96$^{+0.13}_{-0.10}$ & 0.06$^{+0.06}_{-0.03}$  & 1.02$^{+0.13}_{-0.14}$ & 0.02$^{+0.04}_{-0.02}$ & 0.99$^{+0.17}_{-0.12}$ & 0.03$^{+0.09}_{-0.03}$ & 5.8$^{+1.6}_{-1.8}$ & 0.15$^{+0.04}_{-0.05}$ \\
2\arcmin -- 5\arcmin       & 0.86$^{+0.18}_{-0.11}$ & 0.03$^{+0.04}_{-0.02}$  & 0.88$^{+0.17}_{-0.14}$ & 0.00$^{+0.01}_{-...}$  & 0.87$^{+0.18}_{-0.13}$ & 0.06$^{+0.01}_{-0.06}$ & 3.6$^{+1.2}_{-0.8}$ & 0.14$^{+0.05}_{-0.03}$ \\
\multicolumn{9}{c}{\bf A3112}           \\
0\arcmin -- 1.5\arcmin     & 0.81$^{+0.12}_{-0.07}$ & 0.02$^{+0.02}_{-0.02}$  & 1.18$^{+0.11}_{-0.14}$ & 0.08$^{+0.07}_{-0.06}$ & 1.00$^{+0.29}_{-0.26}$ & 0.05$^{+0.10}_{-0.05}$ & 6.3$^{+2.7}_{-1.7}$ & 0.22$^{+0.09}_{-0.06}$ \\
1.5\arcmin --  4.5\arcmin  & 0.85$^{+0.12}_{-0.10}$ & 0.02$^{+0.02}_{-0.02}$  & 0.75$^{+0.10}_{-0.14}$ & 0.00$^{+0.02}_{-...}$  & 0.80$^{+0.17}_{-0.16}$ & 0.01$^{+0.03}_{-0.01}$ & 3.4$^{+0.9}_{-0.7}$ & 0.22$^{+0.06}_{-0.05}$ \\
\multicolumn{9}{l}{\vspace{-0.2cm}} \\
\tableline 
\label{ther.tab}
\end{tabular}
\end{center}
\end{table*}

\section{Soft excess}
\begin{figure*}
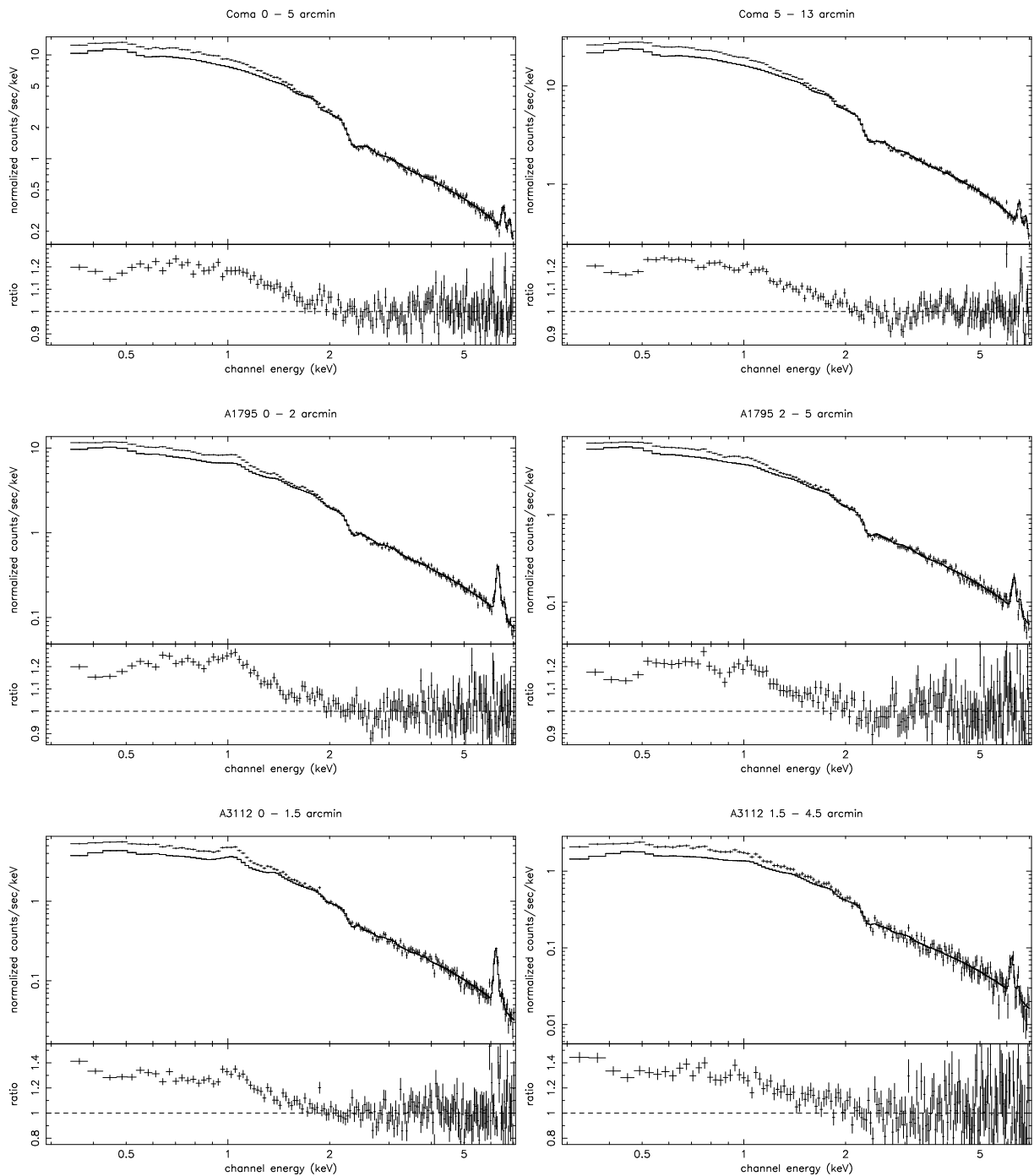

\vbox{
\hbox{
\psfig{figure=f4a.ps,width=8.0cm,angle=-90}
\psfig{figure=f4b.ps,width=8.0cm,angle=-90}}
\vspace{0.5cm}
\hbox{
\psfig{figure=f4c.ps,width=8.0cm,angle=-90}
\psfig{figure=f4d.ps,width=8.0cm,angle=-90}}
\vspace{0.5cm}
\hbox{
\psfig{figure=f4e.ps,width=8.0cm,angle=-90}
\psfig{figure=f4f.ps,width=8.0cm,angle=-90}}}
\vspace{0.5cm}
\caption{The PN data of Coma, A1795 and A3112 with 1$\sigma$ statistical uncertainties. 
The solid line shows the best fit single temperature fit to 2 -- 7 keV data. Lower panels show the ratio of the 
data to the extrapolated model. }
\label{se_pn.fig}
\end{figure*}

\begin{figure*}
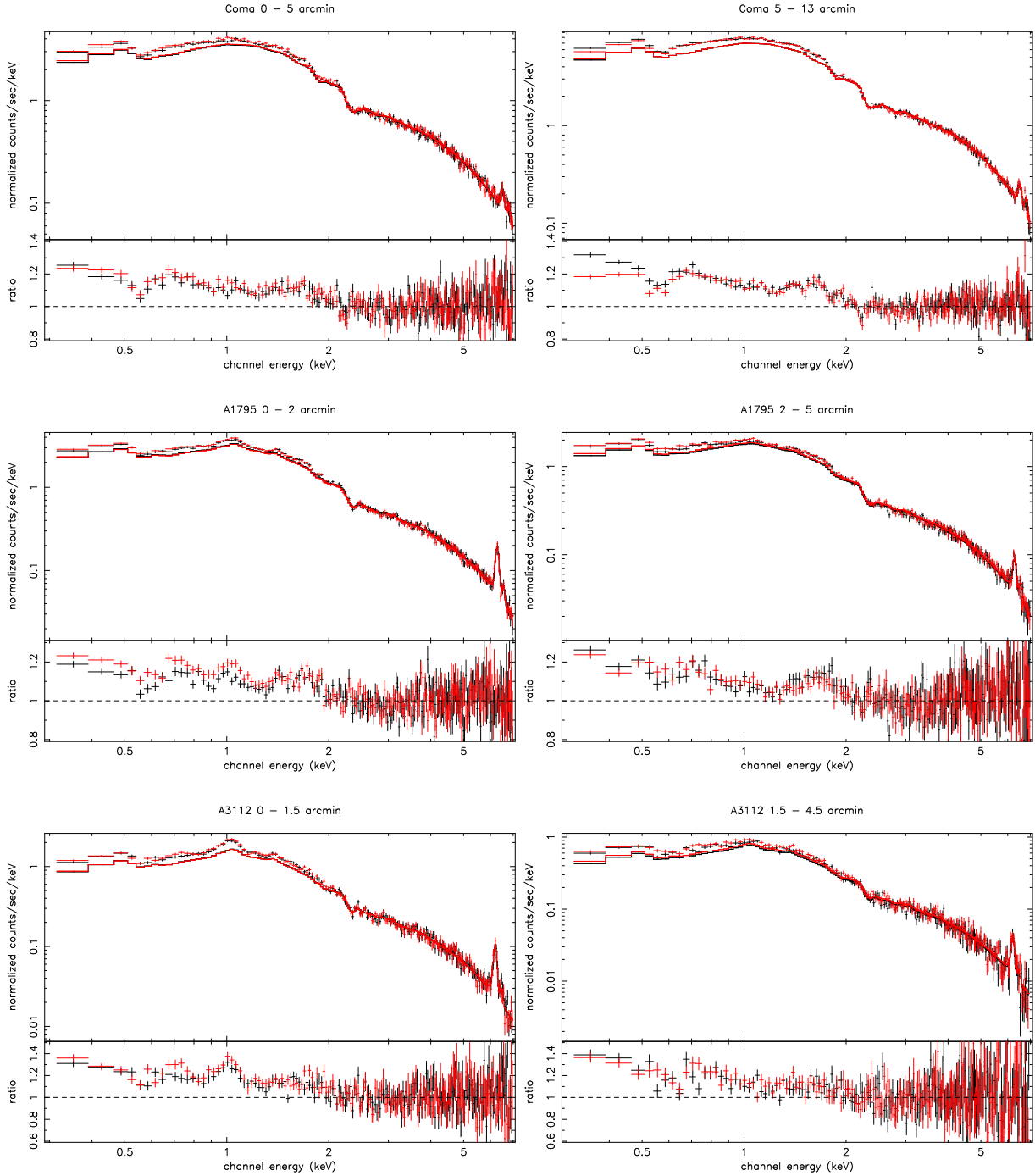

\vbox{
\hbox{
\psfig{figure=f5a.ps,width=8.0cm,angle=-90}
\psfig{figure=f5b.ps,width=8.0cm,angle=-90}}
\vspace{0.5cm}
\hbox{
\psfig{figure=f5c.ps,width=8.0cm,angle=-90}
\psfig{figure=f5d.ps,width=8.0cm,angle=-90}}
\vspace{0.5cm}
\hbox{
\psfig{figure=f5e.ps,width=8.0cm,angle=-90}
\psfig{figure=f5f.ps,width=8.0cm,angle=-90}}}
\vspace{0.5cm}
\caption{Same as Fig. \ref{se_pn.fig}, but for MOS.}
\label{se_m12.fig}
\end{figure*}

Extrapolating the 2 -- 7 keV band thermal models (see \ref{hard}) to soft X-ray energies reveals the soft excess 
(see Figs. \ref{se_pn.fig},\ref{se_m12.fig}) in Coma, A1795 and A3112.
In all cases, both in PN and MOS, the data are above the model at energies below 2 keV. The excess increases towards 
lower energies, reaching 20\% (40\%) of the model level at 0.3 keV for Coma and A1795 (A3112).
The difference of the soft excess fraction in A3112 is a further indication that the effect is not due 
to calibration problems.

\subsection{Thermal modeling}
The residuals above the hot thermal model in the 0.3 -- 2.0 keV band exhibit differences at 5\% level of the model 
between PN and MOS, consistent with the calibration residuals (e.g. Briel 2001). 
This, as a fraction of the soft excess flux maybe large enough to prevent accurate parametrization 
of the soft component.  
We thus proceed by modeling PN and MOS data separately, keeping the hot component parameters fixed to the values
found above, and allowing a second MEKAL emission component to account for 
the soft excess. By comparing the output obtained from PN and MOS data we can estimate the level of 
low energy calibration uncertainties. 
 
The resulting fits (Table \ref{ther.tab}) are statistically acceptable
(see Table \ref{mekal_pow.tab}), with reduced $\chi^{2}$ 
values below unity, indicating that the adopted 5\% systematic error in
the fits is an overestimate of the 
real calibration residuals. The PN and MOS data give consistent
results for the temperature and metal abunbance 
at all cases. Attributing the small offsets in PN and MOS values to systematic errors, rather than random noise,
we take the average values of PN and MOS as best values in each radial bin, and include both PN and MOS statistical 
uncertainty intervals in the final errors of the temperature, metal abundance and the luminosities. Note 
that the normalization differences between PN and MOS are included in
the uncertainties of the luminosities and they dominate 
over the effect of the statistical uncertainties of the parameters of
the very precisely determined hot component, which were 
thus ignored in the reported luminosity values. 

\begin{figure*}
\plotone{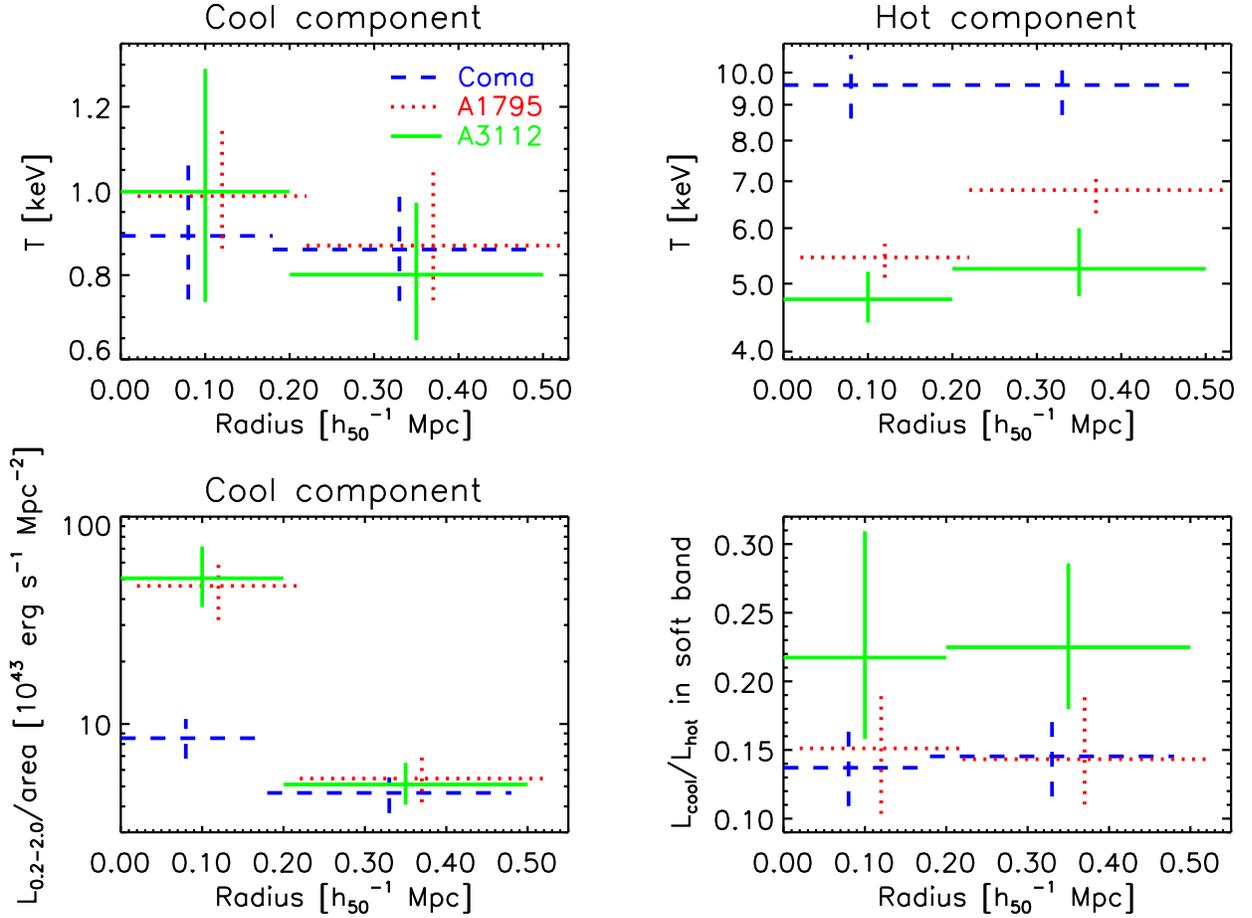}
\caption[]{The best-fit values and 90\% confidence uncertainties for the warm component using a 
thermal model (left panels). Upper right panel shows the temperatures of the hot component.
Blue (dashed), red (dotted) and green (solid) lines linescorrespond to Coma, A1795 and A3113, respectively. The radial bin values (0--0.2--0.5 Mpc)
have been shifted slightly for display purposes. Lower right panel shows the ratio of the luminosities of the hot and 
warm component in 0.2 -- 2.0 keV band.}
\label{soft.fig}
\end{figure*}

The results (see Fig. \ref{soft.fig}) indicate that the soft component has similar temperatures of
0.6 - 1.3 keV in different clusters inside 0.5 h$_{50}^{-1}$ Mpc.
The metal abundances are low, below 0.15 solar within uncertainties
and in most cases consistent with zero. XSPEC simulations indicate that 
the metal abundance of 0.05 of a typical best-fit model can be recovered to a good accuracy, i.e. strong emission 
lines can not be altogether removed by the instrumental redistribution of counts. The 
luminosities of the warm component in 0.2 - 2.0 keV energy band are consistent in different clusters 
in radial range 0.2 - 0.5 h$_{50}^{-1}$ Mpc. In the central 0.2 Mpc the 0.2 - 2.0 keV luminosities are 
consistent within the two cooling flow clusters A1795 and A3112, both being six times higher than in Coma. 
The 0.2 -- 2.0 keV luminosities per metric area of the warm component 
increase by a factor of $\sim 10$ between 0.2 -- 0.5 and 0 - 0.2 Mpc
in A1795 and A3112.
Interestingly, the hot component behaves the same way, producing
constant warm-to-hot component luminosity ratio in a given cluster in  0.2 - 2.0 keV energy band  
0.5 h$_{50}^{-1}$ Mpc, although a variation by a factor of 2 is allowed by the errors.

By holding the temperature of the hot component at the line ratio values, and normalizing the model 
to that of 2 -- 7 keV flux,  one obtains slightly smaller temperatures for the soft component
of Coma, but nevertheless consistent with the ones presented above. 
Further, by letting all the parameters of the hot and warm components vary,
one does not arrive at significant difference in the properties of the soft component from
the ones presented above. 

As an independent check for the existence of the soft excess in these clusters, we 
plotted the ROSAT PSPC data from Bonamente et al. (2002, A3112 not included) together with the best-fit 
two-component PN models (Fig. 7). Due to cross-calibration
uncertainties between PSPC and PN, we allowed a normalization
difference for the model to match the PSPC data at 2 keV
energies. However, the differences were small, a few \% .
The agreement between the XMM-Newton instruments and PSPC is very good in Coma and 
A1795. The conspiracy of similar calibration problems in PN, MOS and PSPC, which incorporate very different
technologies, is extremely unlikely. Thus PSPC data confirm not only the existence but also general spectral 
features of the soft excess. 

\begin{table*}[htb]
\begin{center}
\caption[]{Comparison of best-fits to PN data using mekal or power-law to model the soft excess}
\vspace{0.2cm} 
\begin{tabular}{llllllll}
\tableline
\multicolumn{5}{l}{\vspace{-0.2cm}} \\
          &  \multicolumn{2}{c}{mekal}   &  \multicolumn{2}{c}{power-law}  \\  
          &  \multicolumn{2}{c}{\hrulefill}   &  \multicolumn{2}{c}{\hrulefill}  \\  
radii        &  $\frac{\chi^2}{d.o.f.}$ &  d.o.f.          &   $\frac{\chi^2}{d.o.f.}$    & d.o.f. \\             
\multicolumn{5}{l}{\vspace{-0.2cm}} \\
\multicolumn{5}{c}{\bf Coma}            \\
0\arcmin -- 5\arcmin       & 0.50 & 176 & 0.76 & 177 \\
5\arcmin -- 13\arcmin      & 0.56 & 176 & 0.87 & 177 \\ 
\multicolumn{5}{c}{\bf A1795}                        \\              
0\arcmin -- 2\arcmin       & 0.65 & 177 & 1.08 & 178 \\ 
2\arcmin -- 5\arcmin       & 0.73 & 176 & 0.96 & 177 \\
\multicolumn{5}{c}{\bf A3112}                         \\
0\arcmin -- 1.5\arcmin     & 0.98 & 177 & 1.34 & 178  \\
1.5\arcmin --  4.5\arcmin  & 0.87 & 177 & 1.13 & 178  \\ 
\tableline
\multicolumn{5}{l}{\vspace{-0.2cm}} \\
\label{mekal_pow.tab}
\end{tabular}
\end{center}
\end{table*}

\begin{table*}[b]
\begin{center}
\caption[]{Physical properties of the thermal components. n is the atom density,  $\delta$ is the gas density 
in terms of the critical density. P gives the gas pressure while t$_{cool}$ gives the bremsstrahlung cooling 
time}
\vspace{0.2cm}
\begin{tabular}{lllllllll}
\tableline
\multicolumn{9}{l}{\vspace{-0.2cm}} \\
 & \multicolumn{4}{c}{WARM}  & \multicolumn{4}{c}{HOT}  \\
 & \multicolumn{4}{c}{\hrulefill}  & \multicolumn{4}{c}{\hrulefill}\\
 radii                        &  n$^{a}$  & $\delta$ &  t$_{cool}$   &
P$^{b}$  &  n$^{a}$  & $\delta$ &  t$_{cool}$  &  P$^{b}$  \\
                             &           &          &  (10$^{9}$ y) &
&           &          & (10$^{9}$ y) &                          \\
\multicolumn{9}{c}{\bf Coma}                                       \\
0\arcmin -- 5\arcmin      & 1.5       &  510     & 13  & 2.2 & 3.5 & 1180 & 18 & 55   \\ 
5\arcmin -- 13\arcmin     & 0.6       &  210     & 30  & 0.9 & 1.4 & 460  & 46 & 21   \\
\multicolumn{9}{c}{\bf A1795}                                               \\              
0\arcmin -- 2\arcmin      & 3.2       & 970      & 6   & 5.2 & 7.4 & 2200 & 6 & 65      \\
2\arcmin -- 5\arcmin      & 0.7       & 200      & 27  & 1.0 & 1.6 & 460  & 34 & 17    \\ 
\multicolumn{9}{c}{\bf A3112}                                               \\
0\arcmin -- 1.5\arcmin    & 3.3       & 940      & 6   & 5.6 & 6.2 & 1760 & 7 & 48     \\  
1.5\arcmin --  4.5\arcmin & 0.7       & 190      & 27  & 0.9 & 1.2 & 330  & 40 & 9.6     \\  
\multicolumn{9}{l}{\vspace{-0.2cm}} \\
\tableline
\multicolumn{9}{l}{\vspace{-0.2cm}} \\
\multicolumn{9}{l}{a: [$10^{-3}$ cm$^{-3}$]} \\
\multicolumn{9}{l}{b: [$10^{-12}$ erg cm$^{-3}$]} \\
\end{tabular}
\label{results.tab}
\end{center}
\end{table*}

\subsection{Non-thermal} 
Using a power-law model for the soft component leads to systematically poorer fits 
(see Table \ref{mekal_pow.tab}), but which are statistically acceptable in all cases, except for the center of 
A3112. 
Since most of the reduced $\chi^2$ values for the thermal fits are below unity, this implies an overestimate 
of the systematic errors. If the calibration were better than the 5\% level,  and thus smaller 
systematic error would affect the model, the thermal fits would probably yield  reduced $\chi^2$ values of 
unity, and the power-law fits would all become unacceptable. Thus, although for the moment a firm conclusion of the 
nature of the soft excess is not available, the thermal model is preferred.

\section{Soft component interpretation}

We use the PN and MOS averaged best-fit thermal models obtained above to derive physical properties of the 
warm and hot components (see Table \ref{results.tab}). 
For the estimate of the electron density, we assume that the observed emission measure 
originates from a spherical shell of constant density with radii equal to those of the annuli for which the 
spectra were extracted. The resulting densities of the warm component are similar in different clusters at 
the same radii (2-4 $\times 10^{-3}$ atoms cm$^{-3}$ at 0 - 0.2 h$_{50}^{-1}$ Mpc and 0.6 - 0.7 $\times 10^{-3}$ 
atoms cm$^{-3}$ at 0.2 - 0.5 h$_{50}^{-1}$ Mpc),  
corresponding to overdensities of 1000 - 200 (in terms of the critical density). 

The cooling time scale can be estimated using bremsstrahlung cooling function since the line emission
is negligible due to the apparent low abundances. Using 
\begin{equation}
t_{cool} = {6 \times 10^{9}} ( {{T} \over {10^{6} K}} )^{\frac{1}{2}} ( {{n} \over {10^{-3} cm^{-3}}} )^{-1} years 
\end{equation}
we obtain cooling times larger or comparable to the Hubble time.
The pressure of the ideal (warm) gas, $\sim$10$^{-12}$ erg cm$^{-3}$ is an order of magnitude smaller than 
that of the hot gas.

\section{Conclusions and discussion}
In galaxy clusters Coma, A1795 and A3112, the hot intracluster gas as determined by modeling the  2 -- 7 keV 
PN and MOS data is consistent with that inferred from the FeXXV-FeXXVI line ratio.  This lends confidence to our method 
of characterizing the hot component by using the 2 -- 7 keV band. The expected emission spectrum from this component 
at lower energies may then be computed and compared with the data, during which a significant warm emission component 
at a level above the systematic uncertainties becomes evident.
The non-thermal origin of the phenomenon cannot be ruled 
out at the current level of calibration accuracy, but the thermal
model fits the data better. 
The warm gas is found to have temperatures of 0.6 -- 1.3 keV  inside
0.5 h$_{50}^{-1}$ Mpc, consistent with the 90\% distribution of the
temperature values in Warm Hot Intergalactic Medium WHIM simulations 
(Dave et al., 2001). Within 0.2 and 0.5 h$_{50}^{-1}$ Mpc, the derived
electron densities ($\sim$ 10$^{-4}$ cm$^{-3}$) are marginally
consistent with the simulations, but in the cluster cores the derived 
densities are too high compared to those given by the simulations. This 
indicates that while WHIM may be a viable explanation for the soft excess at 
the outer regions of the clusters, it cannot explain the soft excess 
phenomenon entirely.

While the luminosities of the hot component in Coma, A1795 and A3112  vary substantially, those of the warm component 
outside the cool core region  are consistent in the three clusters being considered.
The similarities suggest a common origin for the warm component, independent of the hot gas and its cooling.
The derived values for the pressure  of the hot component along the central 0.5 Mpc line of sight 
towards Coma, A1795 and A3112 are an  order of magnitude higher than those of the warm gas, suggesting that 
they are not in contact. These requirements can be satisfied  in a WHIM scenario where 
filaments do not penetrate the clusters, but rather form an external network. In this scenario, the density 
of the hot gas in clusters drops faster with radius than that of the WHIM filaments.
Indeed, in Bonamente et al. (2002) the  ROSAT PSPC data pointed to a radial increase of the 
soft excess. Thus, at $\sim$ 1 h$_{50}^{-1}$ Mpc the  pressure equipartition may be attained
and stable structures like filaments may be maintained. Outside the center, the homogenous distribution of filaments, 
projected in the cluster direction, produces constant  luminosity per Mpc$^{2}$ in different clusters. 

One could argue that the central warm component is a result of a cooling flow. However, we note that the central 
brightness peak is not softening; rather the warm and hot component luminosities are enhanced by a similar factor 
compared to the outer parts - this is different  from the standard cooling flow model where the spectral 
peak shifts sharply towards low energies in the center. Also, the cooling flow model does not explain the existence of 
the warm gas outside the cooling radii, as noted in this work. Likewise, the existence of warm gas at the center of Coma 
cannot be explained by cooling, since the temperature of the hot gas is the same inside and outside the core region of 
Coma. On the other hand, we already mentioned that in the center of A1795 and A3112 the hot gas cools and the luminosities 
of the warm and hot gas are enhanced - such phenomena are absent in Coma. This suggests (in the context of the filament
model described above) that 
1) the central line of sight intersects similar amount of WHIM gas in projected filaments, giving a
basic level of radiation ($\sim$ 1 $\times$ 10$^{44}$ erg s$^{-1}$ Mpc$^{-2}$) common to all clusters; 
2) emissions resulting from any cooling of the central hot gas  are reprocessed by the overlying layers of warm gas.

\acknowledgements
J. Nevalainen acknowledges an ESA Research Fellowship, and 
a NASA grant NAG5-9945. M. Bonamente gratefully acknowledges NASA for support.
We thank the referee and Drs. J. Kaastra, M. Markevitch and O. Vilhu for useful comments.

\newpage

\newpage

\newpage

 \newpage

\newpage

\newpage

\newpage

\begin{figure*}
\vbox{
\hbox{
\psfig{figure=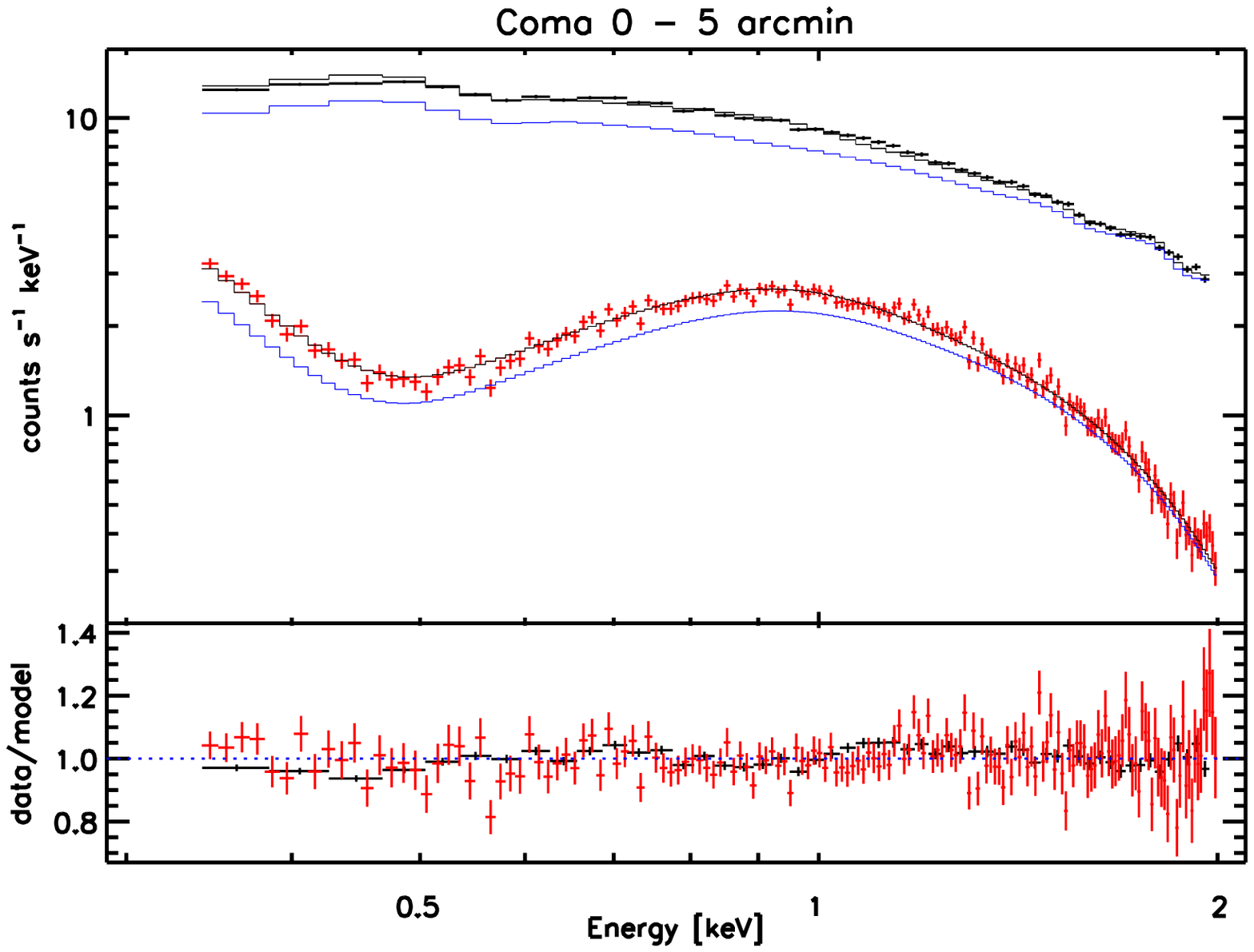,width=8.0cm,angle=0}
\psfig{figure=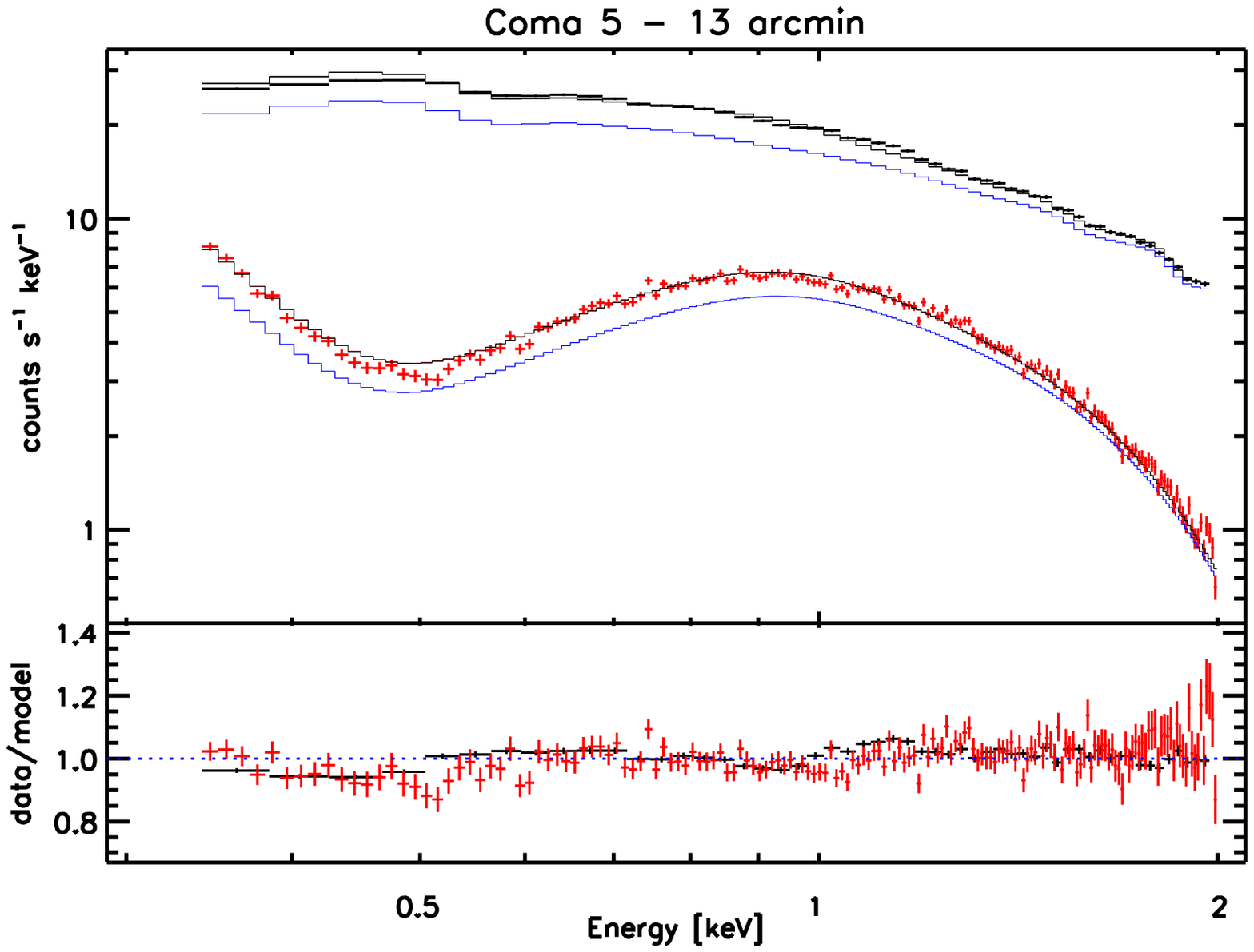,width=8.0cm,angle=0}}
\vspace{0.5cm}
\hbox{
\psfig{figure=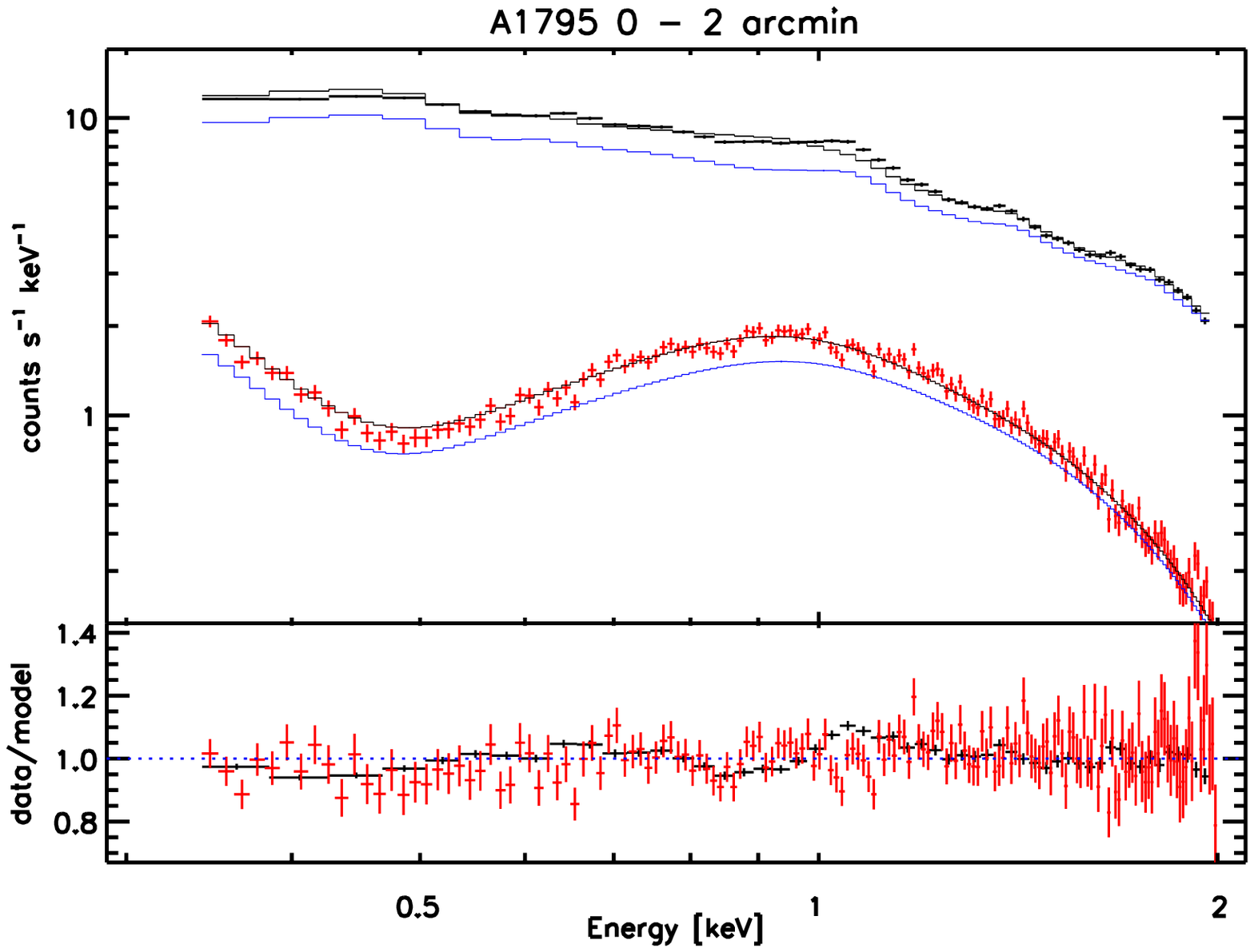,width=8.0cm,angle=0}
\psfig{figure=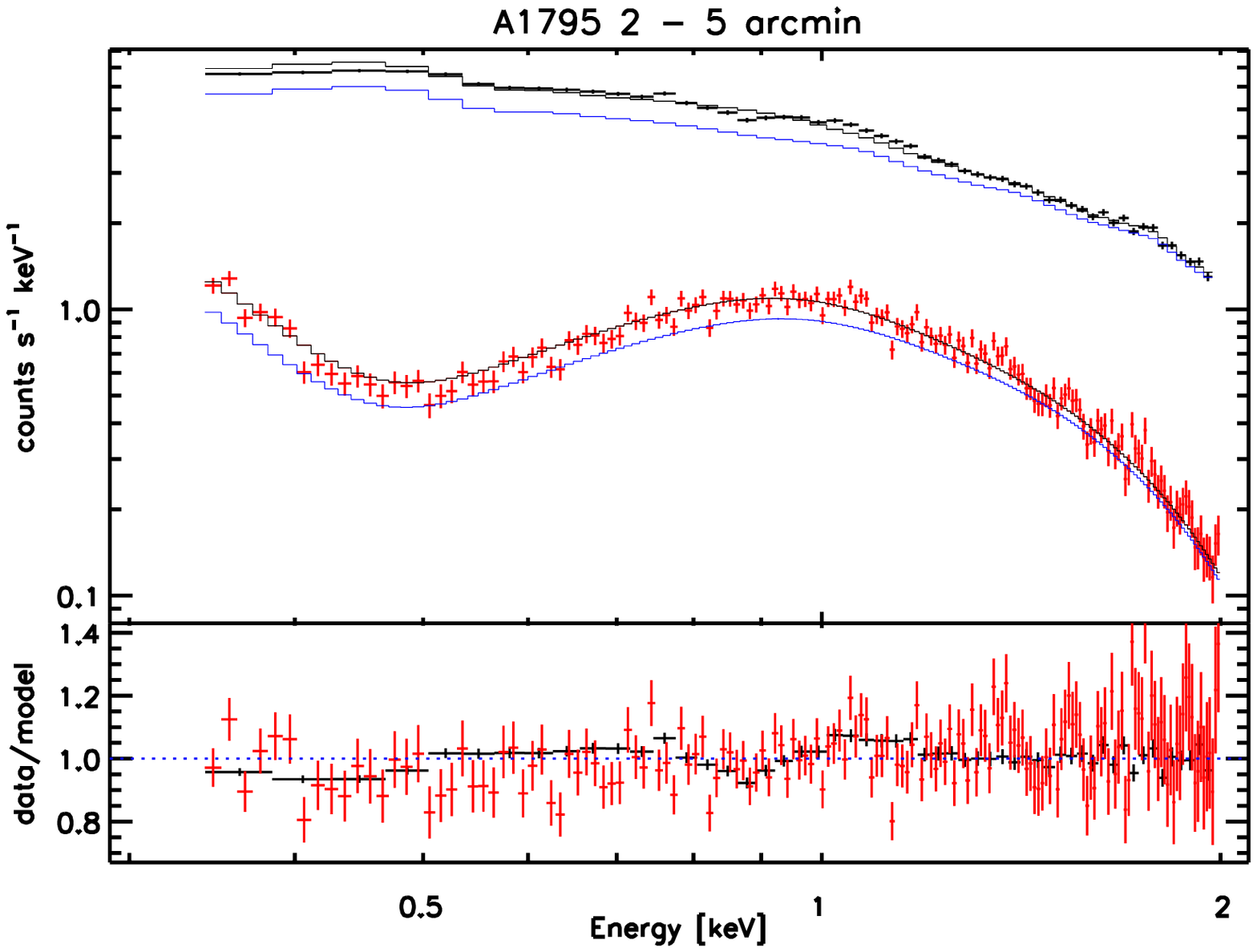,width=8.0cm,angle=0}}}
\vspace{0.5cm}
\caption{The solid black lines show the best fit PN two-component models convolved with the PN and PSPC responses. 
The solid blue lines show the hot gas component separately.
The PN data and the published ROSAT PSPC data (Bonamente et al. 2002)
are shown as black and red crosses, respectively. The lower panel shows the data-to-model 
ratio. The error bars show the  1$\sigma$ statistical uncertainties.}
\label{pp.fig}
\end{figure*}

\end{document}